\documentclass[onecolumn,draftcls]{IEEEtran}
\usepackage{multirow}
\usepackage{amssymb}
\usepackage[cmex10]{amsmath}
\usepackage{graphicx,color}
\usepackage{stfloats}
\title{Pedestrian Motion Direction Estimation Using Simulated Automotive MIMO Radar}
\author{Peter ˜Khomchuk, Inna ˜Stainvas, Igal ˜Bilik
\thanks{}
\thanks{}}
\begin{document}
%  OPTIONAL -->   \ninept            <-- OPTIONAL, for nine pt only
%}
%\ninept
\maketitle

\newcommand{\be}{\begin{equation}}
\newcommand{\ee}{\end{equation}}
\newcommand{\bea}{\begin{eqnarray}}
\newcommand{\eea}{\end{eqnarray}}
\newcommand{\beaa}{\begin{eqnarray*}}
\newcommand{\eeaa}{\end{eqnarray*}}
\newcommand{\ben}{\begin{enumerate}}
\newcommand{\een}{\end{enumerate}}
\newcommand{\bi}{\begin{itemize}}
\newcommand{\ei}{\end{itemize}}
\newcommand{\bs}{\begin{split}}
\newcommand{\es}{\end{split}}

\newcommand{\vecnorm}[1]{\left|\left|#1\right|\right|}

\newcommand{\bml}[1]{\mbox{\boldmath $ #1 $}}
\newcommand{\tbml}[2]{\tilde{\mbox{\boldmath$#1$}}_{\! #2}}
\newcommand{\hbml}[2]{\hat{\mbox{\boldmath$ #1 $}}_{\! #2}}
\newcommand{\bmll}[1]{\mbox{\boldmath $ #1 $}^{(\ell)}}

\newcommand{\one}{\frac{1}{n}}
\newcommand{\half}{\frac{1}{2}\:}
\newcommand{\Dhalf}{\mbox{$\frac{D}{2}$}}
\newcommand{\pomDhalf}{\mbox{$\left(1-\frac{D}{2}\right)$}}
\newcommand{\pxn}{\mbox{$P_{X^n}$}}
\newcommand{\pyn}{\mbox{$P_{Y^n}$}}
\newcommand{\pzn}{\mbox{$P_{Z^n}$}}
\newcommand{\wyxn}{\mbox{$W_{Y^n|X^n}$}}
\newcommand{\qxyn}{\mbox{$Q_{X^nY^n}$}}
\newcommand{\qyxn}{\mbox{$Q_{Y^n|X^n}$}}
\newcommand{\qxyzn}{\mbox{$Q_{X^nY^nZ^n}$}}
\newcommand{\pxyzn}{\mbox{$P_{X^nY^nZ^n}$}}
\newcommand{\pxyn}{\mbox{$P_{X^nY^n}$}}
\newcommand{\pxy}{\mbox{$\bml{P_{XY}}$}}
\newcommand{\qxy}{\mbox{$\bml{Q_{XY}}$}}
\newcommand{\px}{\mbox{$\bml{P_X}$}}
\newcommand{\py}{\mbox{$\bml{P_Y}$}}
\newcommand{\tp}{\mbox{$\tilde{P}$}}
\newcommand{\ptxx}{\mbox{$P_{\tilde{X}X}$}}
\newcommand{\eptxx}{\mbox{${\bf E}_{P_{\tilde{X}X}}$}}
\newcommand{\wyx}{\mbox{$\bml{W_{Y|X}}$}}
\newcommand{\w}{\mbox{$\bml{W}$}}
\newcommand{\qxn}{\mbox{$Q_{X^n}$}}
\newcommand{\qyn}{\mbox{$Q_{Y^n}$}}
\newcommand{\qx}{\mbox{$\bml{Q_X}$}}
\newcommand{\qy}{\mbox{$\bml{Q_Y}$}}
\newcommand{\vxn}{\mbox{$V_{X^n}$}}
\newcommand{\vyn}{\mbox{$V_{Y^n}$}}
\newcommand{\vxyn}{\mbox{$V_{X^nY^n}$}}
\newcommand{\vyxn}{\mbox{$V_{Y^n|X^n}$}}
\newcommand{\x}{\mbox{$\bml{X}$}}
\newcommand{\y}{\mbox{$\bml{Y}$}}
\newcommand{\z}{\mbox{$\bml{Z}$}}
\newcommand{\ixy}{\mbox{$\oo{\bml{I}}(\bml{X};\bml{Y})$}}
\newcommand{\iixy}{\mbox{$\uu{\bml{I}}(\bml{X};\bml{Y})$}}
\newcommand{\ixyzn}{\mbox{$i_{X^nY^n;Z^n}$}}
\newcommand{\ixyznl}{\mbox{$i_{X^nY^n;Z^n}(a^nb^n;c^n)$}}
\newcommand{\ixyznr}{\mbox{$i_{X^nY^n;Z^n}(X^nY^n;Z^n)$}}
\newcommand{\ixxyzn}{\mbox{$i_{X^n;Y^nZ^n}$}}
\newcommand{\ixxyznl}{\mbox{$i_{X^n;Y^nZ^n}(a^n;b^nc^n)$}}
\newcommand{\ixxyznr}{\mbox{$i_{X^n;Y^nZ^n}(X^n;Y^nZ^n)$}}
\newcommand{\iyzn}{\mbox{$i_{Y^n;Z^n}$}}
\newcommand{\iyznl}{\mbox{$i_{Y^n;Z^n}(b^n;c^n)$}}
\newcommand{\iyznr}{\mbox{$i_{Y^n;Z^n}(Y^n;Z^n)$}}
\newcommand{\ixzn}{\mbox{$i_{X^n;Z^n}$}}
\newcommand{\ixznl}{\mbox{$i_{X^n;Z^n}(a^n;c^n)$}}
\newcommand{\ixznr}{\mbox{$i_{X^n;Z^n}(X^n;Z^n)$}}
\newcommand{\vni}{\mbox{$V_{n,i}\left(\left\{y_j^n\right\},c^n\right)$}}
\newcommand{\rni}{\mbox{$R_{n,i}\left(\left\{y_j^n\right\},c^n\right)$}}
\newcommand{\umx}{\mbox{$U_{M_X}\left(\left\{y_j^n\right\},c^n\right)$}}
\newcommand{\tmx}{\mbox{$T_{M_X}\left(\left\{y_j^n\right\},c^n\right)$}}
\newcommand{\tvni}{\mbox{$\tilde{V}_{n,i}(b^n,c^n)$}}
\newcommand{\trni}{\mbox{$\tilde{R}_{n,i}(b^n,c^n)$}}
\newcommand{\tu}{\mbox{$\tilde{U}(b^n,c^n)$}}
\newcommand{\tllt}{\mbox{$\tilde{T}(b^n,c^n)$}}
\newcommand{\sr}{\mbox{$\oo{\bml{R}}$}}
\newcommand{\ir}{\mbox{$\uu{\bml{R}}$}}
\newcommand{\hy}{\mbox{$\oo{\bml{H}}(\bml{Y})$}}
\newcommand{\hx}{\mbox{$\oo{\bml{H}}(\bml{X})$}}
\newcommand{\uh}{\mbox{$\uu{\bml{H}}$}}
\newcommand{\oh}{\mbox{$\oo{\bml{H}}$}}
\newcommand{\er}{\mbox{$\bml{E}$}}
\newcommand{\ron}{\mbox{$\rho_n$}}
\newcommand{\ros}{\mbox{$\rho_s$}}
\newcommand{\ronb}{\mbox{$\bar{\rho}_n$}}
\newcommand{\rob}{\mbox{$\bar{\rho}$}}
\newcommand{\romax}{\mbox{$\rho_{max}$}}
\newcommand{\Pcirc}{\mbox{$\stackrel{\circ}{P}$}}
\newcommand{\Ecirc}{\mbox{$\stackrel{\circ}{E}$}}
\newcommand{\uu}{\underline}
\newcommand{\oo}{\overline}
\newcommand{\reals}{{\rm I\!R}}
\newcommand{\onei}{{\rm 1\!\!\!\:I}}
\newcommand{\dfn}{\stackrel{\triangle}{=}}
\def\ints{\mathop{\rm Z\kern -0.25em Z}\nolimits}
\def\complex{\mathop{|\kern -0.25em \rm C}\nolimits}
\newcommand{\realsd}{\reals^d}
\newcommand{\PP}{\mbox{${\cal P}$}}
\newcommand{\NN}{{\rm I\!\!\!\;N}}
\newcommand{\FF}{\mbox{${\cal F}$}}
\newcommand{\CC}{\mbox{${\cal C}$}}
\newcommand{\LL}{\mbox{${\cal L}$}}
\newcommand{\XX}{\mbox{${\cal X}$}}
\newcommand{\MM}{\mbox{${\cal M}$}}
\newcommand{\AAA}{\mbox{${\cal A}$}}
\newcommand{\DD}{\mbox{${\cal D}$}}
\newcommand{\EE}{\mbox{${\cal E}$}}
\newcommand{\KK}{\mbox{${\cal K}$}}
\newcommand{\DI}{{\DD_{{\rm I}}}}
\newcommand{\Prob}{{\rm Prob\,}}
\newcommand{\la}{\lambda}
\newcommand{\Kalpha}{{K_\alpha}}
\newcommand{\Psialpha}{{\Psi_I(\alpha)}}
\newcommand{\PsiI}{{\Psi_I}}
\newcommand{\calCo}{{\cal{C}}^o}
\newcommand{\calB}{{\cal{B}}}

\newcommand{\Ncal}{{\cal{N}}}

\newcommand{\Shat}{\hat{S}}
\newcommand{\mut}{\tilde{\mu}}
\newcommand{\boldx}{{\bf x}}
\newcommand{\boldX}{{\bf X}}
\newcommand{\boldy}{{\bf y}}
\newcommand{\boldz}{{\bf z}}
\newcommand{\boldp}{{\bf p}}
\newcommand{\uux}{\uu{x}}
\newcommand{\uuY}{\uu{Y}}

\def\muvec{{\mbox{\boldmath $\mu$}}}
\def\Gammavec{{\mbox{\boldmath $\Gamma$}}}
\def\btheta{{\mbox{\boldmath $\theta$}}}
\def\thetat{{\mbox{\scriptsize\boldmath $\theta$}}}

\newcommand{\avec}{{\bf{a}}}
\newcommand{\avect}{{\tilde{\bf{a}}}}
\newcommand{\bvec}{{\bf{b}}}
\newcommand{\cvec}{{\bf{c}}}
\newcommand{\dvec}{{\bf{d}}}
\newcommand{\evec}{{\bf{e}}}
\newcommand{\fvec}{{\bf{f}}}
\newcommand{\epsvec}{{\bf{\epsilon}}}
\newcommand{\pvec}{{\bf{p}}}
\newcommand{\qvec}{{\bf{q}}}
\newcommand{\Yvec}{{\bf{Y}}}
\newcommand{\yvec}{{\bf{y}}}
\newcommand{\uvec}{{\bf{u}}}
\newcommand{\wvec}{{\bf{w}}}
\newcommand{\wvect}{{\tilde{\bf{w}}}}
\newcommand{\xvec}{{\bf{x}}}
\newcommand{\zvec}{{\bf{z}}}
\newcommand{\mvec}{{\bf{m}}}
\newcommand{\nvec}{{\bf{n}}}
\newcommand{\rvec}{{\bf{r}}}
\newcommand{\Svec}{{\bf{S}}}
\newcommand{\Tvec}{{\bf{T}}}
\newcommand{\svec}{{\bf{s}}}
\newcommand{\vvec}{{\bf{v}}}
\newcommand{\gvec}{{\bf{g}}}
\newcommand{\gveca}{\gvec_{\alphavec}}
\newcommand{\uveca}{\uvec_{\alphavec}}
\newcommand{\hvec}{{\bf{h}}}
\newcommand{\ivec}{{\bf{i}}}
\newcommand{\kvec}{{\bf{k}}}
\newcommand{\etavec}{{\bf{\eta}}}
\newcommand{\onevec}{{\bf{1}}}
\newcommand{\zerovec}{{\bf{0}}}
\newcommand{\nuvec}{{\bf{\nu}}}
\newcommand{\alphavec}{{\bf{\alpha}}}
\newcommand{\psivec}{{\bf{\psi}}}
\newcommand{\thetavec}{\boldsymbol{\theta}}
\newcommand{\zetavec}{\boldsymbol{\zeta}}
\newcommand{\phivec}{\boldsymbol{\phi}}
\newcommand{\rhovec}{\boldsymbol{\rho}}
\newcommand{\chivec}{\boldsymbol{\chi}}
\newcommand{\Phivec}{{\bf{\Phi}}}
\newcommand{\Thetavec}{{\bf{\Theta}}}
\newcommand{\deltakvec}{{\bf{\Delta k}}}
\newcommand{\Lambdamat}{{\bf{\Lambda}}}
\newcommand{\invLambdamat}{\Lambdamat^{-1}}
\newcommand{\Gammamat}{{\bf{\Gamma}}}
\newcommand{\Amat}{{\bf{A}}}
\newcommand{\Bmat}{{\bf{B}}}
\newcommand{\Cmat}{{\bf{C}}}
\newcommand{\Dmat}{{\bf{D}}}
\newcommand{\Emat}{{\bf{E}}}
\newcommand{\Fmat}{{\bf{F}}}
\newcommand{\Gmat}{{\bf{G}}}
\newcommand{\Hmat}{{\bf{H}}}
\newcommand{\Jmat}{{\bf{J}}}
\newcommand{\Imat}{{\bf{I}}}
\newcommand{\Kmat}{{\bf{K}}}
\newcommand{\Lmat}{{\bf{L}}}
\newcommand{\Pmat}{{\bf{P}}}
\newcommand{\Pmatperp}{{\bf{P^{\bot}}}}
\newcommand{\Ptmatperp}{{\bf{P_2^{\bot}}}}
\newcommand{\Qmat}{{\bf{Q}}}
\newcommand{\invQmat}{\Qmat^{-1}}
\newcommand{\Smat}{{\bf{S}}}
\newcommand{\Tmat}{{\bf{T}}}
\newcommand{\Tmattilde}{\tilde{\bf{T}}}
\newcommand{\Tmatcheck}{\check{\bf{T}}}
\newcommand{\Tmatbar}{\bar{\bf{T}}}
\newcommand{\Rmat}{{\bf{R}}}
\newcommand{\Umat}{{\bf{U}}}
\newcommand{\Vmat}{{\bf{V}}}
\newcommand{\Wmat}{{\bf{W}}}
\newcommand{\Ymat}{{\bf{Y}}}
\newcommand{\Xmat}{{\bf{X}}}
\newcommand{\Zmat}{{\bf{Z}}}
\newcommand{\Zeromat}{{\bf{0}}}

\newcommand{\Ry}{\Rmat_{\yvec}}
\newcommand{\Rz}{\Rmat_{\zvec}}
\newcommand{\RyInv}{\Rmat_{\yvec}^{-1}}
\newcommand{\Ryhat}{\hat{\Rmat}_{\yvec}}
\newcommand{\Rs}{\Rmat_{\svec}}
\newcommand{\Rn}{\Rmat_{\nvec}}
\newcommand{\Rninv}{\Rmat_{\nvec}^{-1}}
\newcommand{\Reta}{\Rmat_{\etavec}}
\newcommand{\Ralpha}{\Rmat_{\alphavec}}
\newcommand{\Ck}{\Cmat_{\kvec}}
\newcommand{\Cn}{\Cmat_{\nvec}}
\newcommand{\Cg}{\Cmat_{\gvec}}
\newcommand{\invRn}{\Rmat_{\nvec}^{-1}}
\newcommand{\Ical}{{\mathcal{I}}}
\newcommand{\Jcal}{{\mathcal{J}}}
\newcommand{\Rcal}{{\mathcal{R}}}
\newcommand{\Tcal}{{\mathcal{T}}}
% This command splits the parentheses and braces
\newcommand{\LB}{\right. \\  \left.}
\newcommand{\Ccal}{{\mathcal{C}}}
\newcommand{\Fcal}{{\mathcal{F}}}
\newcommand{\LDev}{\right.$ $\left.}
\newcommand{\define}{\stackrel{\triangle}{=}}
%%% For BOLD Greek Letters

\newcommand{\Psimat}{\mbox{\boldmath $\Psi$}}
\newcommand{\bzeta}{\mbox{\boldmath $\zeta$}}
\def\bzeta{{\mbox{\boldmath $\zeta$}}}
\def\btheta{{\mbox{\boldmath $\theta$}}}
\def\bgamma{{\mbox{\boldmath $\gamma$}}}
\def\Beta{{\mbox{\boldmath $\eta$}}}
\def\lam{{\mbox{\boldmath $\Gamma$}}}
\def\bomega{{\mbox{\boldmath $\omega$}}}
\def\bxi{{\mbox{\boldmath $\xi$}}}
\def\brho{{\mbox{\boldmath $\rho$}}}
\def\bmu{{\mbox{\boldmath $\mu$}}}
\def\bnu{{\mbox{\boldmath $\nu$}}}
\def\btau{{\mbox{\boldmath $\tau$}}}
\def\bphi{{\mbox{\boldmath $\phi$}}}
\def\bsigma{{\mbox{\boldmath $\Sigma$}}}
\def\bLambda{{\mbox{\boldmath $\Lambda$}}}
%%% For BOLD Greek Letters
\def\btheta{{\mbox{\boldmath $\theta$}}}
\def\bomega{{\mbox{\boldmath $\omega$}}}
\def\brho{{\mbox{\boldmath $\rho$}}}
\def\bmu{{\mbox{\boldmath $\mu$}}}
\def\bGamma{{\mbox{\boldmath $\Gamma$}}}
\def\bnu{{\mbox{\boldmath $\nu$}}}
\def\btau{{\mbox{\boldmath $\tau$}}}
\def\bphi{{\mbox{\boldmath $\phi$}}}
\def\bPhi{{\mbox{\boldmath $\Phi$}}}
\def\bxi{{\mbox{\boldmath $\xi$}}}
\def\bvarphi{{\mbox{\boldmath $\varphi$}}}
\def\bepsilon{{\mbox{\boldmath $\epsilon$}}}
\def\balpha{{\mbox{\boldmath $\alpha$}}}
\def\bvarepsilon{{\mbox{\boldmath $\varepsilon$}}}
\def\bXsi{{\mbox{\boldmath $\Xi$}}}
\def\betavec{{\mbox{\boldmath $\beta$}}}
\def\betavecsc{{\mbox{\boldmath \tiny $\beta$}}}
\def\xsivec{{\mbox{\boldmath $\xi$}}}
\def\xsivecsc{{\mbox{\boldmath \tiny $\xsivec$}}}
\def\alphavec{{\mbox{\boldmath $\alpha$}}}
\def\alphavecsc{{\mbox{\boldmath \tiny $\alpha$}}}
\def\gammavec{{\mbox{\boldmath $\gamma$}}}
\def\etavecsc{{\mbox{\boldmath \tiny $\eta$}}}
\def\thetavecsc{{\mbox{\boldmath \tiny $\theta$}}}
\def\Ximat{{\mbox{\boldmath $\Xi$}}}

\newcommand{\limn}{\lim_{n \rightarrow \infty}}
\newcommand{\limN}{\lim_{N \rightarrow \infty}}
\newcommand{\limr}{\lim_{r \rightarrow \infty}}
\newcommand{\limd}{\lim_{\delta \rightarrow \infty}}
\newcommand{\phit}{\phi_{\mbox{{\thetat}}}}
\newcommand{\phitk}{\phi_{\mbox{{\thetat}}}}

\newcommand{\AR}
 {\begin{array}[t]{c}
  \longrightarrow \\[-0.3cm]
  \scriptstyle {n\rightarrow \infty}x
  \end{array}}

\newcommand{\RAISE}{{\:\raisebox{.6ex}{$\scriptstyle{>}$}\raisebox{-.3ex}
{$\scriptstyle{\!\!\!\!\!<}\:$}}}

\newcommand{\ARROW}[1]
  {\begin{array}[t]{c}  \longrightarrow \\[-0.4cm] \textstyle{#1} \end{array} }

\newcommand{\pile}[2]
  {\left( \begin{array}{c}  {#1}\\[-0.2cm] {#2} \end{array} \right) }

\newcommand{\ffrac}[2]
  {\left( \frac{#1}{#2} \right)}

\def\squarebox#1{\hbox to #1{\hfill\vbox to #1{\vfill}}}
\newcommand{\qed}{\hspace*{\fill}
            \vbox{\hrule\hbox{\vrule\squarebox{.667em}\vrule}\hrule}\smallskip}

\newcommand{\ifff}{\mbox{\ if and only if\ }}

\newcommand{\ctg}{{\rm ctg}}
\newcommand{\limM}{\lim_{M \rightarrow \infty}}

\newcommand{\eps}{\epsilon}

\newcommand{\degree}{\scriptscriptstyle \circ }

\newcommand{\limsxupn}{\limsup_{n \rightarrow \infty}}
\newcommand{\liminfn}{\liminf_{n \rightarrow \infty}}

\newcommand{\st}{\stackrel}

\newcounter{MYtempeqncnt}

\begin{abstract}
Micro-Doppler-based target classification capabilities of the automotive radars can provide high reliability and short latency to the future active safety automotive features. A large number of pedestrians surrounding vehicle in practical urban scenarios mandate prioritization of their treat level. Classification between relevant pedestrians that cross the street or are within the vehicle path and those that are on the sidewalks and move along the vehicle rout can significantly minimize a number of vehicle-to-pedestrian accidents.

This work proposes a novel technique for a pedestrian direction of motion estimation which treats pedestrians as complex distributed targets and utilizes their micro-Doppler (MD) radar signatures.  
The MD signatures are shown to be indicative of pedestrian direction of motion, and the supervised regression is used to estimate the mapping between the directions of motion and the corresponding MD signatures. In order to achieve higher regression performance, the state of the art sparse dictionary learning based feature extraction algorithm was adopted from the field of computer vision by drawing a parallel between the Doppler effect and the video temporal gradient.

The performance of the proposed approach is evaluated in a practical automotive scenario simulations, where a walking pedestrian is observed by a multiple-input-multiple-output (MIMO) automotive radar with a 2D rectangular array. The simulated data was generated using the statistical Boulic-Thalman human locomotion model. Accurate direction of motion estimation was achieved by using a support vector regression (SVR) and a multilayer perceptron (MLP) based regression algorithms. The results show that the direction estimation error is less than $10^{\circ}$ in $95\%$ of the tested cases, for pedestrian at the range of $100$m from the radar.
\end{abstract}


\section{Introduction}
Autonomous driving is one of the major mega-trends in the automotive industry \cite{Buehler}-\cite{Levinson}. Improving reliability and safety of the current vehicles by enhancing their sensing capabilities (frequently called active safety) is the first step toward autonomous driving. Radar along with LiDAR and vision systems is one of the main automotive sensors \cite{Fleming}-\cite{Murad}. Although current stand-alone automotive radar performance does not meet all sensing requirements, radars are typically included in the majority of automotive active safety systems. Currently, the object detection and localization are the main automotive radar tasks, while the object identification and classification are performed by the vision and LiDAR systems. In order to decrease the latency of the object classification and to enable radar-based systems to operate stand-alone without fusion with other sensing modalities, it is desirable to provide the automotive radars with the target classification capabilities.

Since it is highly important to mitigate vehicle-to-pedestrian accidents, pedestrian recognition is the main classification task of automotive active safety systems. The problem of a binary classification between a vehicle and a pedestrian using data from the automotive radar was successfully addressed in \cite{Heuel1}-\cite{Heuel3} using the state of the art classification techniques like Support Vector Machine (SVM) \cite{Cortes}.

In addition to the target classification, it is desirable to predict trajectories of the surrounding automotive targets. The first step toward this challenging goal is an estimation of the pedestrian motion direction, which can be used to discriminate between pedestrians that are crossing (or intend to cross) the road and those that are walking on the sidewalk along the vehicles motion direction. This work focuses on the problem of a pedestrian direction of motion estimation using an automotive radar.

Most of the automotive targets can be considered as extended targets. In the radar literature extended target is defined as a target that consists of multiple moving parts and occupies multiple spatial cells \cite{Skolnik}. A walking pedestrian observed by the radar with a sufficient spatial resolution can be considered as an extended target. The relative motions of the parts of the extended complex target are called the micro motions \cite{Chen2}. The micro motions generate additional modulations on the radar echo which are typically denoted as the micro-Doppler (MD) effect \cite{Chen1}. Recently it was demonstrated that the MD effect uniquely represents different targets and can be efficiently used for the target classification \cite{KimLing09} - \cite{Bilik07}. Although MD has been widely used for automatic target recognition, to the best of our knowledge this work is the first attempt to apply the MD signatures to a problem of complex target direction of motion estimation.

The MD signature is determined by the radial components of the velocity vectors of the individual scatterers that constitute the target. When the bulk (averaged over all scattering centers) velocity vector changes its direction, the radial components of the velocity vectors of the individual scatterers also change which leads to a change in the resulting MD signature. Thus, the observed MD depends on the direction of the bulk velocity vector which defines the target direction of motion. This work employs a supervised learning approach to estimate the target direction of motion from the observed MD signatures. Two regression methods were used in this work: the Support Vector Regression (SVR) \cite{Smola97, Smola04} and the Multilayer Perceptron (MLP). In contrast to the approach in \cite{Fairchild}, which utilities a radar with two widely separated receivers and relies on the actual estimates of the Doppler shift from an oscillating part of the target to estimate its orientation, the proposed approach observes a complex target from a single angle and extracts the direction of motion information from the entire MD signature. In addition, unlike the tracking algorithm in \cite{Heuel3} we estimate only the instantaneous direction of the target's motion, which can be used as an additional information for the tracking algorithm along with the range and bearing estimates.

The proposed here approach to a direction of motion estimation requires a radar with a high spatial resolution, which is capable of separating different scattering centers of the complex extended target. The MD signatures obtained from multiple spatial micro-cells provide information about the relative positions of the different parts of the target. This additional information is expected to improve the direction of motion estimation. This work considers an automotive MIMO radar with collocated antennas that is capable of providing high azimuth and elevation resolution \cite{LiStoica0}, and transmits linear frequency modulated (LFM) waveforms to achieve high range resolution. The utilization of the MIMO radar is motivated by its ability to achieve high angular resolution using a short sensor array, while utilization of the LFM waveform is motivated by its practical simplicity. Notice that the proposed here approach to direction of motion estimation is not limited to a particular selection of the radar architecture, and is suitable for any radar configuration that is able to provide sufficient spatial resolution. 

Conventionally, application of the supervised learning algorithms to multidimensional data requires a feature extraction (dimensionality reduction) preprocessing. Although in the MD-based target recognition literature multiple feature extraction methods have been studied \cite{KimLing09}-\cite{Bilik07}, this work adopts a sparse-learning-based feature extraction technique originally proposed in the field of computer vision \cite{Sapiro1}-\cite{Sapiro10}. In \cite{Sapiro1}, the sparse-learning-based feature extraction was successfully applied to a problem of video-based classification of human activities. In this work we draw a parallel between the video temporal gradient used in \cite{Sapiro1} and the MD data, and apply the sparse learning approach to reduce dimensionality of the MD signatures of the target. 

The performance of the proposed technique is evaluated via simulations in the scenarios with a walking pedestrian observed by an automotive MIMO radar. The MD signatures of the pedestrian are generated using the Baulic-Thalman human locomotion model \cite{Boulic}. The simulation results show that the accurate direction of motion estimation is possible with low-latency even in relatively low signal-to-noise ratio (SNR) scenarios.

The main novelties of this work are: a) utilization of the MD signatures of the complex extended targets with multiple moving parts to target motion direction estimation; b) application of the supervised regression algorithms to the problem of motion direction estimation; c) adaptation of the computer-vision-based feature extraction method used for human activities classification to the radar MD-based motion direction estimation; d) numerical study of the proposed direction of motion estimation approach in the automotive scenarios with walking pedestrian; and e) numerical evaluation of the various MIMO radar configurations in terms of the motion direction estimation performance.

The rest of the paper is organized as follows. Section II states a received signal model for the collocated MIMO radar which observes extended target with multiple moving parts. Section III describes a scenario of a pedestrian motion direction estimation and corresponding choice of the automotive radar parameters. Sparse modeling and feature extraction methods are discussed in Section IV. Section V evaluates the performance of the proposed direction of motion estimation approach via numerical simulations considering a scenario of a pedestrian direction of motion estimation using an automotive radar. Finally, our conclusions are summarized in Section VI.

\section{Radar Signal Model}
\label{SignalModel}
\begin{figure}[thb]
\centerline{\includegraphics[scale=0.6]{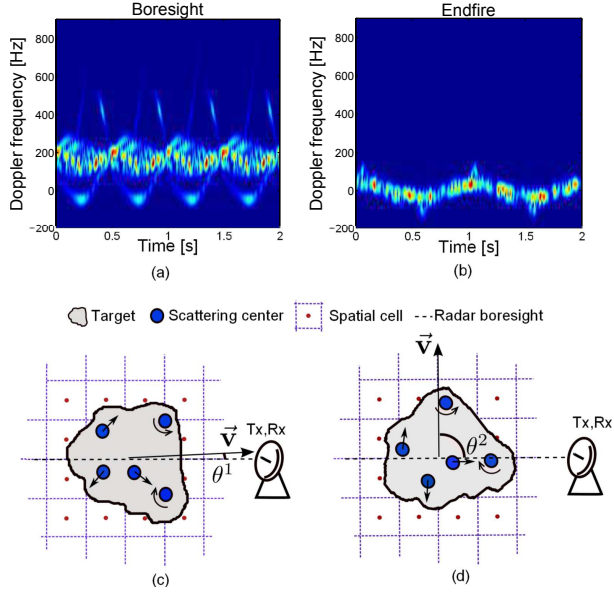}}
%\centerline{\psfig{figure=Fig1.eps,width=3.4in,height=3.4in}}
\caption{The MD signature as a function of the target direction of motion. Subplots (a) and (b) show the spectrograms obtained from the simulated data for a pedestrian walking along the LOS and perpendicular to the LOS, respectively. The radar data is simulated using the Boulic-Thalman human locomotion model \cite{Boulic}. Subplots (c) and (d) schematically show arbitrary extended target moving with the two different directions of motion observed by the monostatic radar. The extended target consists of multipe scattering centers which in addition to the bulk target's velocity $\vec{\vvec}$ perform different types of micro-motions. As a result of different target directions of motion $\theta^1$ and $\theta^2$ the radar observes different MD signatures. For example, the signatures in subplots (a) and (b) could correspond to the targets directions $\theta^1$ and $\theta^2$. In addition, for different directions of motion the radar observes different relative positions of the scattering centers.}\label{setup}
\end{figure}

In the monostatic radar scenarios the MD effect depends only on the radial velocities of the individual parts of the target. A change in the direction of the target's bulk velocity vector results in changes in the radial velocities of the individual parts of the target and leads to changes in the observed MD effect. Hence, the direction of the target's motion relatively to the radar defines the target's MD signature. For example, a comparison between Fig. \ref{setup}(a) and Fig. \ref{setup}(b) shows that the MD signatures of a pedestrian walking along the Line-of-sight (LOS) with the radar and perpendicular to the LOS are significantly different. 

The target direction of motion with respect to the vehicle is defined as the angle, $\theta$, between the target's bulk motion direction and the boresight of the radar, as shown in Fig. \ref{setup}(c) and Fig. \ref{setup}(d). Notice that the angle, $\theta$, describes the general direction of motion of the complex target rather than of its individual scatterers. This work proposes a supervised regression method for motion direction estimation, $\theta$, of the complex target based on its MD signature.

This section develops a radar signal model for the MD signatures of the complex target obtained by a MIMO radar with $M_t$ transmitting elements and $M_r$ receiving elements antenna arrays \cite{LiStoica1}. High angular resolution provided by the MIMO radar is required to separate groups of individual scatterers within the same extended target, and to obtain more information about their relative locations and motions. This additional information is required to achieve reasonable direction of motion estimation performance.

Consider an automotive MIMO radar that observes an extended complex target which consists of the $Q$ independent scattering centers. Fig. \ref{setup}(c) and Fig. \ref{setup}(d) show that the location, $\uvec_q(\theta)$, and the velocity, $\vvec_q(\theta)$, of the $q$th, $q=1,\ldots,Q$, individual scattering center depend on the target direction of motion $\theta$. For the clarity of presentation, the dependence of the locations and velocities of the scattering centers on the target direction of motion is explicitly denoted as follows: $\uvec_q(\theta) = \uvec_q^{\theta}$ and $\vvec_q(\theta) = \vvec_q^{\theta}$.

Let $\avec(\uvec_q^{\theta}) = [a_1(\uvec_q^{\theta}),a_2(\uvec_q^{\theta}),\ldots,a_{M_t}(\uvec_q^{\theta})]^T$ and $\bvec(\uvec_q^{\theta}) = [b_1(\uvec_q^{\theta}),b_2(\uvec_q^{\theta}),\ldots,b_{M_r}(\uvec_q^{\theta})]^T$ be a $M_t \times 1$ transmitting and a $M_r \times 1$ receiving array response vectors to a scattering center located at $\uvec_q^{\theta}$, respectively. Let the $k$th, $k=1,2,\ldots,M_t$ antenna element of the transmitting array transmit a sequence of $P$ narrowband finite-duration pulses of the waveform $s_k(t)$ at the pulse repetition frequency (PRF) $f_r = 1 / T_r$, where $T_r$ is a temporal duration of the transmitted pulse. The baseband radar echo received at the $l$th antenna element of the receiving array due to the transmission from the $k$th antenna element of the transmitting array and scattering from the $Q$ scattering centers is
\bea
\label{Model1}
r_{kl}(t, \theta) & = & \sum_{q=1}^Q\eta_q b_l(\uvec_q^{\theta}) a_k(\uvec_q^{\theta}) \\
& \times & \sum_{p=0}^{P-1}s_k(t-pT_r-\tau(\uvec_q^{\theta}))e^{j 2 \pi f_d\left(\vvec_q^{\theta}\right)pT_r} + e_{kl}(t) \nonumber
\eea
where $\eta_q$, $\tau(\uvec_q^{\theta})$ and $f_d(\vvec_q^{\theta})$ are a complex reflection coefficient of the $q$th scattering center, a round-trip time-delay, and the Doppler-shift induced by the $q$th scattering center respectively, and $e_{kl}(t)$ is a spatio-temporal additive complex zero-mean white circular Gaussian noise with a variance $\sigma^2 \delta(\tau)$. Assuming that the scattering centers are independent and that the corresponding complex reflection coefficient $\eta_q$, $q=1,2,\ldots,Q$ remain constant from pulse-to-pulse, the coherent processing interval (CPI) can be defined as $T_c = PT_r$ \cite{Skolnik}. The variable $p$ in (\ref{Model1}) is often referred as a slow-time \cite{Richards2}, and the Doppler shift $f_d(\vvec_q^{\theta})$, which is assumed to be constant during the CPI, is called a slow-time Doppler shift.

The following assumption on the orthogonality of transmitted waveforms simplifies the joint processing of the received signals transmitted by the different transmitting antennas.

\textit{Orthogonal waveforms:}  Assuming pulses of the duration $T$, the waveforms transmitted by the $i,j=1,\ldots,M_t$ transmitters are orthogonal if
\be
\label{orthog1}
\int_T s_i(t)s_j^*(t) dt = \left\lbrace
\begin{array}{cc}
\int_T\vert s_i(t) \vert ^ 2 dt = 1, & i=j \\
0, & i \neq j
\end{array} \right.
\ee
where $(^*)$ denotes the complex conjugate operator.

Orthogonality of the transmitted waveforms provides separability of the received signals, thus creating in total $M_t \times M_r$ independent and separable transmitter-target-receiver paths. Arranging $M_t \times M_r$ received signals modeled as in (\ref{Model1}) in one $ M_tM_r \times 1$ vector, the received signal at the time $t$ can be modeled as
\be
\rvec(t, \theta) = \sum_{q=1}^Q\eta_q\bvec(\uvec_q^{\theta}) \otimes \left( \avec(\uvec_q^{\theta}) \odot \muvec(t,\tau(\uvec_q^{\theta}), f_d(\vvec_q^{\theta})) \right) + \evec(t)
\label{Model3}
\ee
where $\muvec(t,\tau(\uvec_q^{\theta}), f_d(\vvec_q^{\theta}))$ is the $M_t \times 1$ vector with the $k$th element of the following form
\be
\mu_k(t,\tau(\uvec_q^{\theta}), f_d(\vvec_q^{\theta})) = \sum_{p=0}^{P-1}s_k(t-pT_r-\tau(\uvec_q^{\theta}))e^{j 2\pi pT_rf_d(\vvec_q^{\theta})} \nonumber
\ee
and $\otimes$ and $\odot$ denote the Kroneker and the Hadamard products, respectively.

Let the target be placed on the three dimensional spatial grid of $N$ non-overlapping cells with the cell centers located at $\tilde{\uvec}_i$, $i=1,2,\ldots,N$ and the center of the grid at every time instance $t$ coinciding with the geometric center of the target. Such a spatial grid can be defined by a bank of spatial filters
\be
\label{grid}
\Gmat = \left[\gvec_1,\gvec_2,\ldots,\gvec_N\right]
\ee
where $\gvec_i = \bvec(\tilde{\uvec}_i)\otimes\left(\wvec\odot\avec(\tilde{\uvec}_i)\muvec(t,\tau(\tilde{\uvec}_i),0)\right)$ is a spatial filter matched to the center of the $i$th spatial cell. The range-gated and the beamformed signal at the slow-time $p$ from the cell $i$ can be written as follows \cite{LiStoica0}
\bea
\label{Model4}
x_{ip}(\theta) & = & \int_{t=pT_r}^{( p+1 ) T_r}\gvec_i^H( t )\rvec( t, \theta ) dt  \\
%& = & \sum_{q=1}^Q \eta_q\bvec^H(\tilde{\uvec}_i)\bvec(\uvec_q) \int_{t=pT_r}^{( p+1 ) T_r}  \left(\wvec \odot \avec(\tilde{\uvec}_i)\odot\muvec(t,\tau(\tilde{\uvec}_i),0)\right)^H (\wvec\odot\avec(\uvec_q)\odot\muvec(t,\tau(\uvec_q), f_d(\uvec_q))dt + n_i(p) \\
& = & \sum_{q=1}^{Q}\gamma_{iq}(\theta)e^{j2\pi f_d(\vvec_q^{\theta})pT_r} + n_i(p) \nonumber
\eea
where
\bea
\gamma_{iq}(\theta) & = & \eta_q \bvec^H(\tilde{\uvec}_i)\bvec(\uvec_q^{\theta}) \sum_{k=1}^{M_t}a_k^*(\tilde{\uvec}_i)a_k(\uvec_q^{\theta})  \nonumber \\
& \times & \int_{t=0}^{T_r} s_k(t - \tau(\tilde{\uvec}_i))^*s_k(t - \tau(\uvec_q^{\theta}))dt \nonumber
\eea
is an amplitude of the radar echo received from the scattering center $q$ after range-gating and beamforming by the spatial filter $\gvec_i$, and
\be
n_i(p) = \int_{t=pT_r}^{( p+1 ) T_r}\gvec_i^H( t )\evec( t ) dt \nonumber
\ee
is a white zero-mean complex Gaussian process with variance $\sigma_n^2 = \sigma^2M_tM_r$ uncorrelated for different spatial cells if the centers of the corresponding spatial cells are sufficiently separated. In the matrix form, (\ref{Model4}) can be rewritten as follows
\be
\label{MatrixModel}
\xvec_i(\theta) = \Hmat_i(\theta) \gammavec_i(\theta) +\nvec_i
\ee
where $\xvec_i(\theta)$ is the $P\times 1$ slow-time Doppler signal received from the spatial cell with the center at $\tilde{\uvec}_i$, $\gammavec_i(\theta) = [\gamma_{i1}(\theta), \gamma_{i2}(\theta),\ldots,\gamma_{iQ}(\theta)]^T$ is a vector of corresponding amplitudes, $\nvec_i = [n_{1}, n_{2},\ldots,n_{P}]^T$ is a noise vector $\nvec_i \sim C\Ncal(\zerovec,\sigma_n^2\Imat_{P\times P})$, and $\Hmat(\theta)=[\hvec_{1}(\theta),\hvec_{2}(\theta),\ldots,\hvec_{Q}(\theta)]$ is a $P\times Q$ matrix, with the following slow-time temporal steering vectors in its columns
\bea
\hvec_q(\theta) = \left[ e^{j2\pi f_d(\vvec_q^{\theta})1T_r}, e^{j2\pi f_d(\vvec_q^{\theta})2T_r}, \right.\\ 
\left. \ldots, e^{j2\pi f_d(\vvec_q^{\theta})PT_r}\right]^T, q = 1,\ldots,Q \nonumber
\eea
Notice that the slow-time radar echo in (\ref{MatrixModel}) received from the spatial cell $i$ explicitly depends on the direction of motion of the extended target, $\theta$. For the performance evaluation, the SNR can be define as the ratio of the signal power averaged over the spatial grid cells in (\ref{grid}) to the noise power
\be
\label{SNR}
SNR(\theta) = \frac{1}{N}\frac{\sum_{i=1}^N\lVert\xvec_i(\theta)\rVert_2^2}{\sigma_n^2}
\ee

\section{System Parameters}
\label{SysPrms}
This section provides the reasoning for the selection of the specific transmitter and receiver array configurations, waveform type, and parameters of the spatial filter bank, required for sufficiently informative MD signatures obtained from the received radar echo in (\ref{MatrixModel}).

\begin{figure}[thb]
\centerline{\includegraphics[scale=0.75]{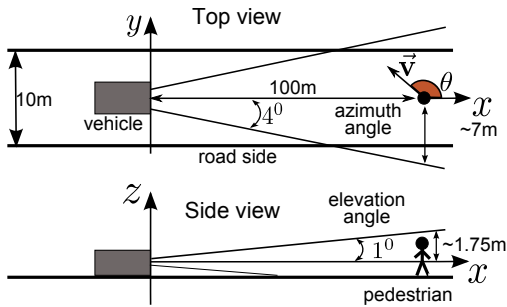}}
%\centerline{\psfig{figure=Fig2.eps,width=2.5in,height=1.5in}}
\caption{Walking pedestrian observed by the automotive radar.}\label{Fig1}
\end{figure}

Consider an automotive scenario in Fig. \ref{Fig1}. The road is located in the $xy$ plane such that the $x$ axis is pointing along the road. In the considered here simulation scenarios, the elevation of the terrain was assumed to be constant and the target motion was assumed to be solely in a 2D range-azimuth ($xy$) plain. For convenience we also introduce a spherical coordinate system, such that a point in space is defined by the vector $\uvec = [r,\beta,\gamma]^T$, where $r$, $\beta$ and $\gamma$ are the range, the azimuth and the elevation, respectively. Notice that in Fig. \ref{Fig1}, the range axis and the $x$ axis coincide. Fig. \ref{Fig1} shows a static vehicle equipped with an antenna array located at the origin $\uvec_a = [0, 0, 0]^T$ observing a walking pedestrian. At the time instance $t=0$, the pedestrian is located at $l_p=100$m away from the vehicle at the coordinates $\uvec_p = [l_p, 0, 0]^T$. The pedestrian's direction of motion, $\theta$, is defined with respect to the $x$ (range) axis. 

Typically a long range automotive radar needs to operate only in the narrow azimuth and elevation field-of-view (FOV). In the simulated scenarios, the road is assumed to be $D_r = 10$m wide, which at the distance of $l_p = 100$m from the vehicle results in the maximum azimuth angle of $\lvert\beta_{max}\rvert = \tan^{-1}\frac{D_r}{2l_p}  < 4^{\circ}$. Similarly, the elevation angle is limited by the height of the pedestrian. Assuming maximum height of the pedestrian to be $h_{max} = 2$m, at the distance of $l_p = 100$m from the radar, maximum elevation angle is approximately $\lvert\gamma_{max}\rvert = \tan^{-1}\frac{D_r}{2l_p}< 1^{\circ}$. The values of the range $r$ are also limited, since the pedestrian, who is detected $l_p = 100$m away from the radar, cannot significantly change its position during the observation time.

\begin{figure}[htb]
\centerline{\includegraphics[scale=0.5]{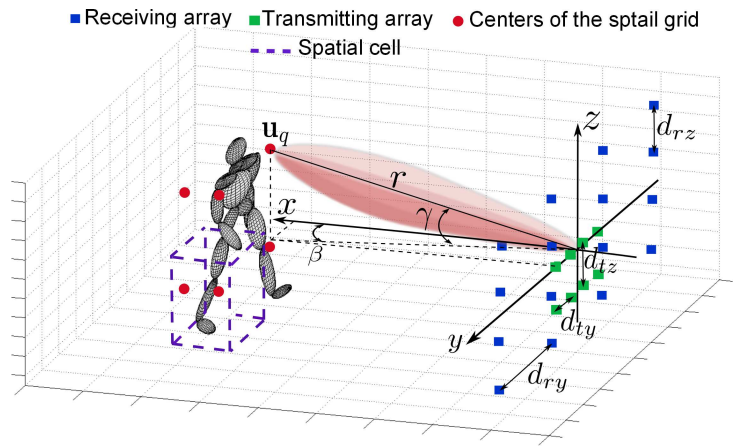}}
%\centerline{\psfig{figure=Fig3.eps,width=3.1in,height=1.9in}}
\caption{Walking pedestrian observed by the antenna array.}\label{Fig2}
\end{figure}

The vehicle is assumed to be equipped with a 2D transmitting and receiving antenna arrays (Fig. \ref{Fig2}), operating at the frequency of $f_c = 24$GHz. Both arrays are located in the $zy$ plane and have the same phase center. Let the receiver be an $M_r = L_{ry}\times L_{rz}$ uniform rectangular array (URA) with $L_{ry}$ elements spaced by $d_{ry}$ in each row, and $L_{rz}$ elements spaced by $d_{rz}$ in each column. Similarly, let the transmitter be an $M_t = L_{ty} \times L_{tz}$ URA, with the corresponding element spacing $d_{ty}$ and $d_{tz}$. The transmitting and receiving array responses to the target located at $\uvec_q$ can be defined as follows
\bea
\label{steeringvecs}
\avec(\uvec) & = & \left[e^{j2\pi\phi_{t1}(\uvec)} e^{j2\pi\phi_{t2}(\uvec)} \ldots e^{j2\pi\phi_{tM_t}(\uvec)} \right]^T \nonumber\\
\bvec(\uvec) & = & \left[e^{j2\pi\phi_{r1}(\uvec)} e^{j2\pi\phi_{r2}(\uvec)} \ldots e^{j2\pi\phi_{rM_r}(\uvec)} \right]^T\nonumber
\eea
\bea
\label{PhaseShifts}
\phi_{tk}(\uvec_q) & = & \sin \beta_{q} \cos\gamma_q\frac{d_{ty}}{\lambda} \left(i_{ty}-\frac{L_{ty}-1}{2}\right) \nonumber\\
& + & \sin \gamma_{q}\frac{d_{tz}}{\lambda}\left(i_{tz}-\frac{L_{tz}-1}{2}\right), \nonumber \\
i_{ty} & = & 1, \ldots,L_{ty}, i_{tz} = 1,\ldots,L_{tz} \nonumber\\
\phi_{rl}(\uvec_q) & = & \sin \beta_{q} \cos\gamma_q\frac{d_{ry}}{\lambda} \left(i_{ry}-\frac{L_{ry}-1}{2}\right) \nonumber\\
& + & \sin \gamma_{q}\frac{d_{rz}}{\lambda} \left(l_{rz}-\frac{L_{rz}-1}{2}\right), \nonumber \\
i_{ry} & = & 1,\ldots,L_{ry}, i_{rz} = 1,\ldots,L_{rz} \nonumber
\eea
where $\lambda$ is a wavelengths, and $k = i_{tz} L_{ty} + i_{ty}$, $l = i_{rz} L_{ry} + i_{ry}$.

The radar spatial resolution defines dimensions of the cells in Fig. \ref{Fig2} and the spatial grid given in (\ref{grid}). Since pedestrian's torso, arms and legs have different motion characteristics, the ability to resolve radar echoes received from the different body parts is important to obtain more information about pedestrian's motion, and thus achieve sufficient regression performance. However, such approach requires small spatial cells and hence high spatial resolution. 

The following subsections discuss the transmitting and receiving array configurations and the transmitted waveform parameters required to achieve sufficient azimuth, elevation, range and Doppler resolutions.

\subsection{Azimuth and Elevation Resolution}
\label{subsec:AzimElev}
This work considers a MIMO radar that achieves sufficiently high angular resolution \cite{LiStoica0}. The MIMO radar with colocated antennas creates a so-called \textit{virtual} aperture, which is equal to the convolution of the transmitting and receiving array apertures \cite{LiStoica1}. Hence, for a MIMO radar with $M_t$ transmitting and $M_r$ receiving elements the resulting aperture consists of the $M_tM_r$ \textit{virtual} antenna elements.

\begin{figure}[htb]
\centerline{\includegraphics[scale=0.5]{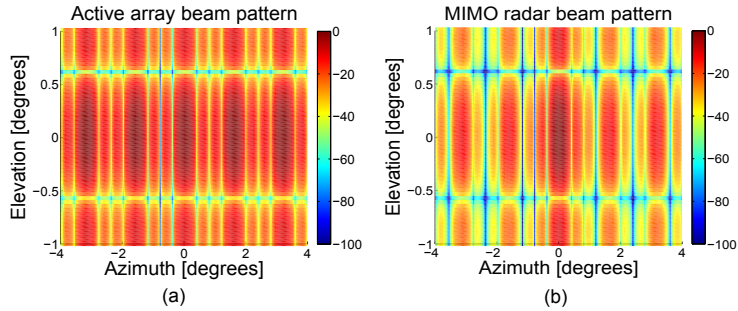}}
%\centerline{\psfig{figure=Fig4.eps,width=3.2in,height=1.5in}}
\caption{Beam patterns of: (a) SIMO radar with a single antenna transmitter and $L_{ry} = 4$ by $L_{rz} = 3$ receiving array with $d_{ry} = 36\lambda$ and $d_{rz} = 32\lambda$; (b) MIMO radar with $L_{ty} = 4$ by $L_{tz} = 1$ transmitting array with inter element spacings $d_{ty} = 12\lambda$, and $L_{ry} = 4$ by $L_{rz} = 3$ receiving array with $d_{ry} = 36\lambda$ and $d_{rz} = 32\lambda$.} \label{Fig3}
\end{figure}

Fig. \ref{Fig1} shows that for practical automotive scenarios the FOV between $-4^{\circ}$ and $4^{\circ}$ in the azimuth direction and between $-1^{\circ}$ and $1^{\circ}$ in the elevation direction is sufficient to cover the entire FOV of interest. The grating lobes outside the region of interest are acceptable, and therefore the antenna elements can be separated by more than $\lambda/2$. Fig. \ref{Fig3}(a) shows the beam pattern of the SIMO radar with $L_{ry} = 4$ by $L_{rz}=3$ rectangular receiving antenna array and the interelement spacings $d_{ry} = 36\lambda$ and $d_{rz} = 32\lambda$ in the horizontal and vertical directions, respectively. Notice the grating lobes in the FOV of interest. Suppression of the grating lobes in this SIMO radar requires more dense receiving array. Alternatively, Fig. \ref{Fig3}(b) shows the beam pattern of the MIMO radar with the same receiving array as in the SIMO case, but with $L_{ty} = 4$ by $L_{tz} = 1$ transmitting array with the interelement spacings $d_{ty} = 12\lambda$. Notice, that there are no grating lobes in the FOV of interest. Thus, the MIMO radar provides the desirable beam pattern in the FOV of interest using a smaller number of antennas compared to the SIMO system.

The elevation and the azimuth dimensions of the spatial grid in (\ref{grid}) are determined by the beamwidth of the MIMO radar. From the beam pattern in Fig. \ref{Fig3}(b) the half power beamwidth in the azimuth and the elevation directions is $\triangle \beta = 0.39^{\circ}$ and $\triangle \gamma = 0.6^{\circ}$, respectively.

The MIMO radar with transmitting array of $0.45$m and 2-D receiving array of $1.35$m $\times$ $0.9$m was considered here to fit into the dimensions of a conventional vehicle. Optimization of the automotive MIMO radar configuration is a subject of our future work.

\subsection{Range and Doppler Resolution}
\label{subsec:RangeDoppler}
Let the $(k_{y},k_{z})$th antenna element of the transmitting array transmit a sequence of LFM chirps
\be
\label{lfm}
s_k(t) = e^{j\pi\frac{f_B}{T_0}\left(t - \frac{1}{2}T_0\right)^2}\left[h(t)-h(t - T_0)\right]
\ee
where $f_B$ is the bandwidth of the chirp, $T_0$ is the pulse duration, and $h(t)$ is the Heaviside step function. Bandwidth of the transmitted LFM waveform defines the range resolution $\triangle r = c / (2f_B)$, and therefore the range dimension of the spatial cells in the grid given by (\ref{grid}). Following FCC regulations, the bandwidth of $f_B = 250$MHz was used for the LFM radar at $24$GHz, which results in a range resolution of $\triangle r = 0.6$m.

Let the maximum velocity of the pedestrian's body parts be $v_{max} = 3$ m/s, which at the radar carrier frequency of $f_c = 24$GHz generates a Doppler shift of $f_{D_{max}} = 2 v_{max} f_c / c = 480$Hz. Since the MD signatures in (\ref{MatrixModel}) are obtained from the slow-time data, maximum observed Doppler shift defines the pulse repetition period, which in order to avoid aliasing was set to be $T_r = 1 / (2f_{D_{max}}) \approx 1$ms.

The Doppler frequency resolution $\triangle f$ is defined by the smallest change in the target's velocity $\triangle v$ that needs to be identified. Let $\triangle v = 0.2m/s$, which results in the Doppler resolution of $\triangle f = 32$Hz. Since the dwell duration defines the Doppler resolution, the number of transmitted LFM chirps was set to be $P = 32 \geq 1 / (\triangle f T_r)$. The LFM chirp duration was set to be $T_0 = 1 \mu$s.

Each antenna element of the MIMO transmitting array transmits an orthogonal waveform. Assuming that the LFM chirps in (\ref{lfm}), transmitted by the different transmitting antenna elements, have the same bandwidth and duration, the orthogonality assumption in (\ref{orthog1}) can be achieved by the time-division multiplexing or frequency division. Notice that the proposed direction of motion estimation approach is not limited to the LFM waveforms chosen here for the practical simplicity, and can be used with any other waveforms.

\subsection{Spatial grid}
\label{spatial_grid}
Typically, the automotive radar first detects the moving target and then estimates its location. The estimated location is used to center the spatial grid in (\ref{grid}) on the pedestrian target. According to the model in (\ref{MatrixModel}), the radar echo is received from the center of each cell. Fig. \ref{Fig2} shows the 3-D spatial grid superimposed on the pedestrian target. The dimensions of the grid cells are defined by the array and waveform parameters. Since we are interested in resolving parts of the human body located in the adjacent spatial cells, the dimensions of the cells can be chosen according to the half power beamwidths and the range resolution values. Hence, the cell size in the spherical coordinates is selected to be $\triangle \uvec = [ \triangle r, \triangle \beta, \triangle \gamma]^T = [0.6, 0.39, 0.6]^T$, which when converted to the Cartesian coordinates at the range of $l_p = 100$m becomes $[0.6, 0.68, 1.04]^T$m. Furthermore, the pedestrian is always assumed to be located inside the $2 \times 2 \times 2$ spatial grid which results in $N=8$ spatial cells. The locations of the centers of the cells with respect to the center of the grid can be found from the following Cartesian product
\be
\triangle \rvec \times \triangle \betavec \times \triangle \gammavec \nonumber
\ee
where $\triangle \rvec = [-\triangle r, \triangle r]^T$, $\triangle \betavec = [-\triangle \beta, \triangle \beta]^T$, and $\triangle \gammavec = [-\triangle \gamma, \triangle \gamma]^T$. Notice that the spatial grid is assumed to consist of only the relevant spatial cells i.e. the spatial cells where the target (pedestrian) is located.

\section{Direction of Motion Estimation}
\begin{figure}[htb]
\centerline{\includegraphics[scale=0.9]{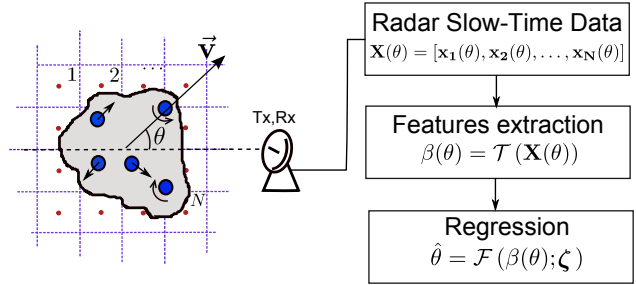}}
%\centerline{\psfig{figure=Fig5.eps,width=3in,height=1.3in}}
\caption{Regression-based target direction of motion estimation approach.} \label{Regr}
\end{figure}

The model in (\ref{MatrixModel}) characterizes a complex extended target by a single parameter $\theta$ - the target's motion direction. Fig. \ref{Regr} schematically shows the regression-based target direction of motion estimation approach. The key point of this approach is to find a parametric function $\Fcal$, which for a given set of radar echoes $\xvec_i(\theta)$, $i=1,2,\ldots,N$, received from the complex target with unknown $\theta$ provides the following mapping:
\bea
\label{GeneralRegression}
\hat{\theta} & = & \Fcal\left(\betavec(\theta); \zetavec\right) \nonumber \\
\betavec(\theta) & = & \Tcal\left(\xvec_1(\theta),\xvec_2(\theta),\ldots,\xvec_N(\theta)\right) \nonumber
\eea
where $\hat{\theta}$ is an estimate of the true direction of motion $\theta$, $\zetavec$ is a vector of regression parameters estimated using a database of \textit{a-priori} collected radar echoes received from the complex target with known directions of motion, and $\Tcal$ is a dimensionality reduction or feature extraction transformation applied to the raw radar data. This section discusses the steps of the proposed regression-based target direction of motion estimation approach shown in Fig. \ref{Regr}.

\subsection{Data Generation}
\label{LocoModel}
The first step of the proposed in Fig. \ref{Regr} direction of motion estimation approach is collection of radar echoes which contain the MD signatures of the target of interest. The MD radar echoes $\xvec_i(\theta)$ in (\ref{MatrixModel}) received from the walking pedestrian can be synthesized using a human locomotion model. This work adopts the Boulic-Thalman model from \cite{Boulic}, and its implementation from \cite{Bradley} and \cite{Chen11}. The Boulic-Thalman model is based on the empirical mathematical parametrization applied to a biomechanical experimental data in order to obtain an averaged human walking model which does not contain any information about personalized motion features. A walking human is represented as a stickman with $17$ characteristic points, e.g. knees, elbows, thorax. The model provides 3D positions of the segments of the body defined by these points as a function of time. In total, the motion is described by $12$ trajectories, $3$ translations and $14$ rotations. These trajectories, translations and rotations describe one cycle of a human body motion - a period between two successive contacts of the left heel with the floor. The cycle is defined by a relative velocity and height of the human.

The outputs of the Boulic-Thalman model are the time-varying locations of the $Q = 17$ characteristic points. These locations are used to calculate the locations $\uvec_{q}$, $q=1,\ldots,Q$ of the $Q$ body parts. The velocities of the body parts are obtained as the rate of change of the corresponding locations. Each body part is assumed to be an independent elliptical scattering center (Fig. \ref{Fig2}) with the reflection coefficient calculated using the radar cross section (RCS) of the ellipsoid \cite{Trott}. The radar echoes from the walking pedestrian are generated using the obtained locations, velocities and reflection coefficients in (\ref{Model3}). The slow-time radar echoes in (\ref{MatrixModel}) received from multiple spatial cells are obtained using the spatial grid in (\ref{grid}).

The Boulic-Thalman model is parametrized by the walking velocity $v_b$ and the body height $h_b$ of the pedestrian. In order to make the simulated data more realistic the following distortion factors were introduced: a) a randomly time-varying $h_b$ was uniformly distributed between $1.6m$ to $2m$; b) a time-varying $v_b$ due to the random acceleration distributed normally with the zero mean and the standard deviation $0.008m/s^2$ (initial velocity was $v_b = 1m/s$); and c) the normally time-varying motion direction $\theta$ with the mean value at the true angle and the standard deviation $0.03$ radians.

\begin{figure}[htb]
\centerline{\includegraphics[scale=0.5]{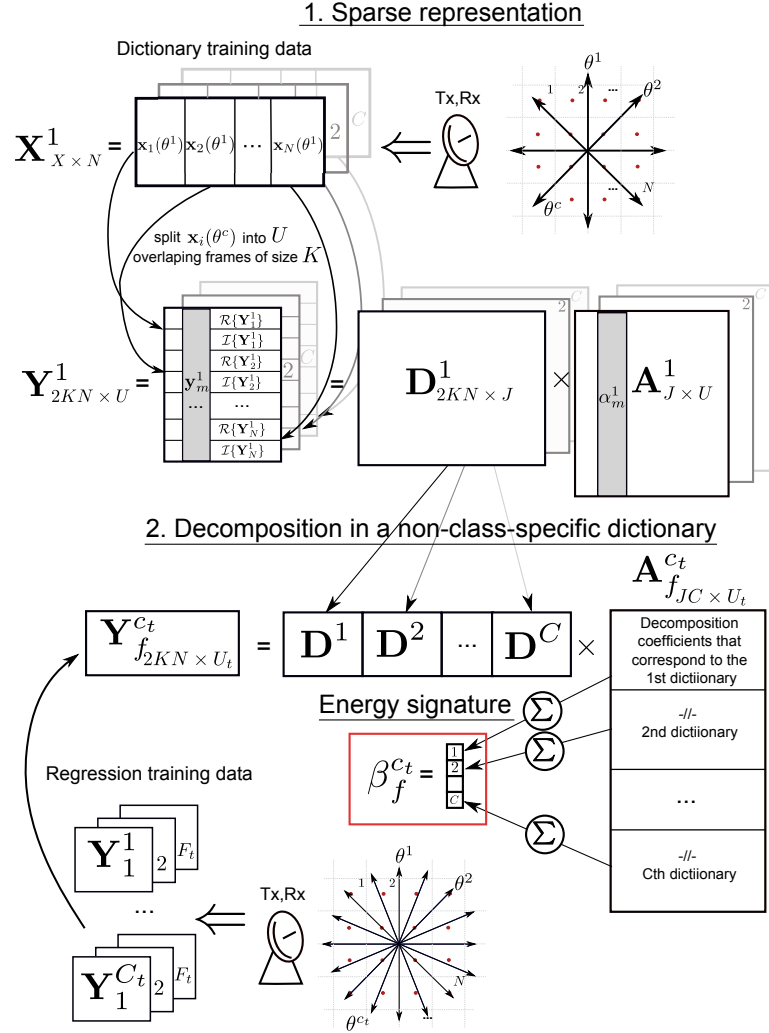}}
%\centerline{\psfig{figure=Fig6.eps,width=3.4in,height=4.5in}}
\caption{Sparse dictionary learning-based feature extraction. First, a collection of slow-time radar echoes $\Xmat^c$ received from the target moving with direction $\theta^c$, $c=1,\ldots,C$ is reshaped with overlap into a matrix $\Ymat^c$ of data samples (see Fig. \ref{Split}). Obtained data samples are then used to learn a sparse dictionary $\Dmat^c$ and the corresponding decomposition coefficients $\Amat^c$. This procedure is repeated for $C$ basic directions of motion and the obtained dictionaries are combiend into a non-class-specifict dictionary $\Dmat$. Further, data from the larger set of directions of motions $\theta^{c_t}$, $c_t=1,\ldots,C_t$ is decomposed in $\Dmat$. The obtained decomposition coefficients $\Amat_f^{c_t}$ are used to calculate the $C\times 1$ energy signatures $\beta_f^{c_t}$ by summing the absolute values of the decomposition coefficients which corresponds to the same basic directions of motions. The obtained energy signatures are further used to train the regression model for the direction of motion estimation.} \label{FeEx}
\end{figure}

\subsection{Feature Extraction}
\label{SparseModeling}
Since the estimation of the pedestrian motion signatures requires target observations over a considerably long time-period, the dimensionality of the radar echos $\xvec_i(\theta)$ becomes large. Processing high-dimensional data is computationally intensive and requires a large training database (curse of dimensionality problem \cite{Bellman}), thus the data dimensionality reduction or the feature extraction $\Tcal$ is typically used prior to application of the regression algorithms.

Various feature types for the problem of MD-based target recognition were proposed in the literature during last decade \cite{KimLing09}-\cite{Bilik07}.
Good classification results were demonstrated using physical model-based features \cite{Bilik07}, information theoretic features \cite{Tekeli}, speech processing motivated features (cepstrum, mel-frequency cepstrum (mfcc), linear predictive coding (lpc)) \cite{Bilik06, Bilik12}, and others \cite{Bar}. However, selection of an optimal feature set for MD-based target classification remains an open research question.

This work adopts the feature extraction approach proposed in \cite{Sapiro1}, where the sparse dictionary learning was used to classify human activities via video temporal gradient. The video temporal gradient captures differences between the two consecutive video frames. In this problem, the video temporal gradient is analogous to the Doppler frequency shift, since for the short time interval the Doppler shift is linearly proportional to the relative changes in the target's position. 

The sparse dictionary learning-based feature extraction reduces the data dimensionality to a small number, $C$, of basic target directions of motion, whose combination is used to represent all other possible directions. Thus, the proposed direction of motion estimation process can be presented as a two-stage approach in Fig. \ref{FeEx}. In the first stage, the set of the $C$ sparse dictionaries is learned from the training data. In the second stage, any radar measurement that strongly depends on the target direction of motion is decomposed in these dictionaries. The rest of this section describes each stage in details.

\begin{figure}[htb]
\centerline{\includegraphics[scale=0.6]{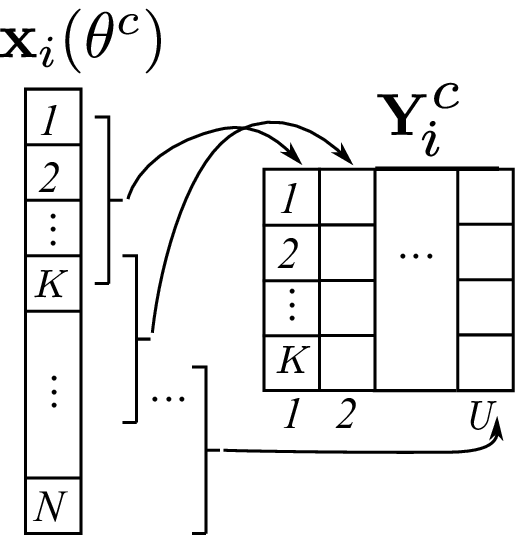}}
%\centerline{\psfig{figure=Fig7.eps,width=1.3in,height=1.2in}}
\caption{Splitting slow-time radar echoes $\xvec_i(\theta^c)$ into $U$ overlaping frames of the size $K$ to form a data sample matrix $\Ymat_i^c$.} \label{Split}
\end{figure}

\subsubsection{Stage 1: Dictionary learning}
Let $\Lambda = \lbrace (\Xmat^1,\theta^1), (\Xmat^2, \theta^2),\ldots, (\Xmat^{C},\theta^{C})\rbrace$ be a dictionary training data set, where an $X \times N$ matrix $\Xmat^c = \left[ \xvec_1(\theta^c),\xvec_2(\theta^c),\ldots,\xvec_N(\theta^c) \right]$ is the collection of the $X \times 1$ slow-time radar echos in (\ref{MatrixModel}) received from $N$ spatial cells when observing the target moving at direction $\theta^c$, $c=1,\ldots,C$. Each column of $\Xmat^c$ (each slow-time signal $\xvec_i(\theta^c)$ ) is split as shown in Fig. \ref{Split} into $U$ overlapping frames of the size $K$, thus forming the $K \times U$ data sample matrices\footnote{The optimal overlap percent can vary depending on a particular scenario, and can be determined using a cross-validation procedure.} $\Ymat_i^c$, $i=1,\ldots,N$. The training data for the $c$th dictionary contains the radar echoes obtained from all spatial cells of interest (cells that contain the target) and has the following form
\bea
\label{dicttraindata}
\Ymat_{2KN \times U}^c = \left[\Rcal\lbrace\Ymat_1^c\rbrace; \Ical\lbrace\Ymat_1^c\rbrace;  \Rcal\lbrace\Ymat_2^c\rbrace; \Ical\lbrace\Ymat_2^c\rbrace; \right. \\ 
\ldots,\left. \Rcal\lbrace\Ymat_N^c\rbrace; \Ical\lbrace\Ymat_N^c\rbrace\right] \nonumber
\eea
where $\Rcal\lbrace\cdot\rbrace$ and $\Ical\lbrace\cdot\rbrace$ denote the real and the imaginary parts of the argument. Each column vector $\yvec_m^c$, $\forall m = 1,\ldots, U$ of the matrix $\Ymat^c$ (the $m$th training sample for the dictionary $c$) consists of the radar echoes received from the $N$ spatial cells of interest when observing the target moving with $c$th basic direction, thereby adding spatial information about the observed extended target to the training data.

The column vectors in $\Ymat^c$ can be represented using the following linear model
\be
\label{SparseModel}
\yvec_m^c = \Dmat^c \alphavec_m^c + \nvec_m^c\;
\ee
where $\nvec_m^c$ is the $2KN\times 1$ additive noise vector with the limited energy, $\lVert \nvec_m^c \rVert_2^2 < \epsilon$, $\Dmat^c$ is the $2KN \times J$ possibly overcomplete ($J > 2KN$) dictionary with $J$ atoms, and $\alphavec_m^c$ is the $J \times 1$ sparse vector of coefficients indicating atoms of $\Dmat^c$ that represent data vector $\yvec_m^c$. The dictionary $\Dmat^c$ and the corresponding vectors of the sparse coefficients $\alphavec^c_m$, $m=1,\ldots,U$ can be learned from the training data by solving the following optimization problem
\be
\label{Dictionary Learning}
\left(\hat{\Dmat}^c,\hat{\Amat}^c\right) = arg \min_{\Dmat^c, \Amat^c}\frac{1}{2}\lVert\Dmat^c\Amat^c-\Ymat^c\rVert_F^2 + \xi \sum_{m=1}^U\lVert\alphavec_m^c\rVert_1\;
\ee
where $\lVert \cdot \rVert_F^2$ is the matrix Frobenius norm, and the $J \times U$ matrix $\Amat^c = \left[\alphavec_1^c, \alphavec_2^c,\ldots,\alphavec_U^c\right]$ contains the sparse decomposition coefficients of the columns of the training data matrix $\Ymat^c$. Minimization of the first summand in (\ref{Dictionary Learning}) decreases the error between the original data and its representation, while minimization of the second summand preserves the sparsity of the obtained solution. The coefficient $\xi$ controls the trade-off between the reconstruction error and the sparsity. The optimization problem in (\ref{Dictionary Learning}) can be numerically solved using modern convex optimization techniques, and this work uses the SPArse Modeling Software (SPAMS) toolbox \cite{Sapiro09, Sapiro10}.

MD signatures for different target's motion directions have similarities, therefore following the approach proposed in \cite{Sapiro1}, we construct the following non-class-specific dictionary which contains characteristics of the $C$ basic directions
\be
\label{bigdictionary}
\Dmat_{2KN\times JC} = [\Dmat^1, \Dmat^2,\ldots,\Dmat^{C}]
\ee 
According to this approach every measurement is represented as the combination of the selected basic directions of motion, while the corresponding decomposition coefficients are used as the features for classification or regression. Therefore, it is desirable for the learned dictionaries to represent as many data variations as possible. Notice that, the angles represented in the data set $\Lambda$ do not have to be uniformly spaced. For example, the pedestrian motion directions where the MD signatures are weak ($\theta$ close to $90^{\circ}$ and $270^{\circ}$) could be represented using more training data. The selection of the optimal basic directions of motions is a topic of our future research.

\subsubsection{Stage 2: Signature vectors}
The constructed dictionary $\Dmat$ obtained in the Stage 1 is now used for the feature extraction. Let $\Lambda_t = \lbrace (\Xmat_1^1,\theta^1),\ldots, (\Xmat_{F_t}^1,\theta^1), \ldots, (\Xmat_1^{C_t}, \theta^{C_t}),\ldots, (\Xmat_{F_t}^{C_t},\theta^{C_t})\rbrace$ be a regression training data set, where each one of the $F_t$ data blocks $\Xmat_f^{c_t}$, $f=1,\ldots,F_t$ is an $X_t \times N$ matrix that contains slow-time radar echoes received from the $N$ spatial cells while observing a target moving at direction $\theta^{c_t}$, $c_t = 1,\ldots,C_t$. 

Time $T_F$ defines the target observation period required for the decision on the target motion direction. Let the pulse repetition period be $T_r$, then the target observation time $T_F$ and the dimensionality of the regression training data vector $X_t$ are related as $X_t=T_F/T_r$. In order to represent more directions of motion in the regression training data without increasing the number of dictionaries, we assume that $\Lambda_t$ contains the radar data from a larger number of different directions than $\Lambda$ (i.e. $\Lambda \in \Lambda_t$ ).

Each of the $N$ columns of $\Xmat_f^{c_t}$ is split into $U_t$ overlapping frames of the size $K$ to form $K \times U_t$ matrices $\Ymat_{fi}^{c_t}$, $i=1,\ldots,N$. Similarly to (\ref{dicttraindata}), these matrices are combined into a ${2KN \times U_t}$ sample matrix, $\Ymat_f^{c_t}$. The columns of $\Ymat_f^{c_t}$ can be represented using the dictionary $\Dmat$ by solving the following convex optimization problem
\be
\label{Decomposition}
\hat{\Amat}_f^{c_t} = arg \min_{\Amat_f^{c_t}} \frac{1}{2} \lVert \Dmat\Amat_f^{c_t} - \Ymat_f^{c_t} \rVert_F^2 + \xi \sum_{j=1}^{U_t} \lVert \alphavec_{fj}^{c_t} \rVert_1
\ee
where $\Amat_f^{c_t} = [\alphavec_{f1}^{c_t},\alphavec_{f2}^{c_t},\ldots \alphavec_{fU_t}^{c_t}]$ is a ${JC \times U_t}$ matrix of the corresponding sparse decompositions. The $JC \times 1$ vector $\alphavec_{fj}^c = [(\alphavec_{fj}^{c_t})_1,\ldots,(\alphavec_{fj}^{c_t})_J,\ldots,(\alphavec_{fj}^{c_t})_{JC}]^T$, which is the sparse representation of the $j$th data sample from $\Ymat_f^{c_t}$ in the merged dictionary $\Dmat$, contains the decomposition coefficients of the $c_t$th target's direction in the basis constructed from the $C$ basic directions. The contribution of the $c$th basic direction to the decomposition of the data matrix $\Ymat_f^{c_t}$ can be obtained by the following summation of the absolute values of all decomposition coefficients that correspond to the basic direction $c$ over $U_t$ data samples
\be
\label{signatures}
(\betavec_f^{c_t})_c = \sum_{j=1}^{U_t}\sum_{i=(c-1)J+1}^{cJ}\lvert (\alphavec_{fj}^{c_t})_i \rvert^2
\ee
The vector $\betavec_f^{c_t} = [(\betavec_f^{c_t})_1, (\betavec_f^{c_t})_2,\ldots,(\betavec_f^{c_t})_C]^T$ can be considered as the energy signature of the data samples $\Ymat_f^{c_t}$, where each entry of the $\betavec_f^{c_t}$ represents the energy contributed by the corresponding basic direction of motion. Using the signature vectors as features reduces the dimensionality of the data from $X_t$ to the number of basic directions $C$. In addition, the signature vectors capture information about relations between different directions of motion. Notice that the summation in (\ref{signatures}) over relatively small number of samples $U_t$ in $\Ymat_f^{c_t}$ is expected to provide significantly higher robustness of the energy signature. Using the signature vectors extracted from $F_t$ training data blocks for each of the $C_t$ different directions, the following regression training data set can be constructed: $\Gamma_{t} = \lbrace (\Bmat^1, \theta^1),(\Bmat^2, \theta^2),\ldots,(\Bmat^{C_t},\theta^{C_t})\rbrace$, where $\Bmat^{c_t} = [\betavec_1^{c_t},\betavec_2^{c_t},\ldots,\betavec_{F_t}^{c_t}]$, $c_t=1,\ldots,C_t$. Notice that the described here sparse-learning-based feature extraction from the radar MD data and the proposed usage of the energy signatures are in general applicable to a variety of multispectral data-based classification problems, such as target's motion direction estimation, pedestrian activities classification, and ground moving target recognition.

\subsection{Regression}
\label{Regression}

The regression training data set $\Gamma_{t}$ can be used to estimate the mapping $\Fcal(\betavec(\theta); \zetavec)$ between the feature vectors $\betavec_j^{c_t}$ and the corresponding direction of motion $\theta^{c_t}$. The mapping function $\Fcal(\betavec(\theta); \zetavec)$ is called a regression model. The feature vector $\betavec(\theta)$, extracted from the radar echoes $\Xmat(\theta)$ that are not represented in the training database, can be used to predict the unknown direction $\theta$. There are multiple methods that can be used for the regression model learning, and this work uses the two common method: SVR \cite{Smola97, Smola04} and MLP-based regression \cite{Hornik89}. 

The performance of the chosen regression model can be evaluated using the testing data set $\Gamma_{s} = \left\{ (\Bmat^1, \theta^1),(\Bmat^2, \theta^2),\right.$ $\ldots,\left.(\Bmat^{C_s},\theta^{C_s})\right\}$, where $\Bmat^{c_s}$ is a $C \times F_s$ matrix which contains $F_s$ feature vectors that correspond to the direction angle $\theta^{c_s}$, $c_s=1,\ldots,C_s$. Note that $\Gamma_s$ contains feature vectors from the directions that are not represented in the training data set $\Gamma_t$. In this work the mean squared error (MSE) between the true directions and the directions predicted by the regression algorithm is used as the quantitative performance metric
\be
\label{MSE}
MSE^{c_s} = \frac{1}{F_s}\sum_{j=1}^{F_s}\left[\theta^{c_s} - \Fcal(\betavec_j^{c_s})\right]^2
\ee
where $F_s$ is the number of available testing radar echoes received from the target moving in direction $\theta^{c_s}$, and the mapping function $\Fcal$ is obtained by the SVR or MLP algorithm.

\section{Simulation Results}
\label{NumericalRes}
This section evaluates performance of the proposed direction of motion estimation approach using MD signatures of a walking pedestrian generated from the Boulic-Thalman model in the scenario described in Section \ref{SysPrms}. First, the regression error of the proposed direction of motion estimation approach is analyzed as a function of the SNR and the observation time $T_F$. Then the direction of motion estimation performance is compared for different radar configurations which result in different number of spatial cells. Finally, in order to provide more insight about the obtained results, the probability of a target direction of motion estimation error being less than a given value is evaluated. The slow-time radar data (\ref{MatrixModel}) have been obtained using the set of spatial filters in (\ref{grid}), and the corresponding feature vectors were estimated using the sparse modeling approach discussed in Section \ref{SparseModeling}.

\begin{figure}[htb]
\centerline{\includegraphics[scale=0.6]{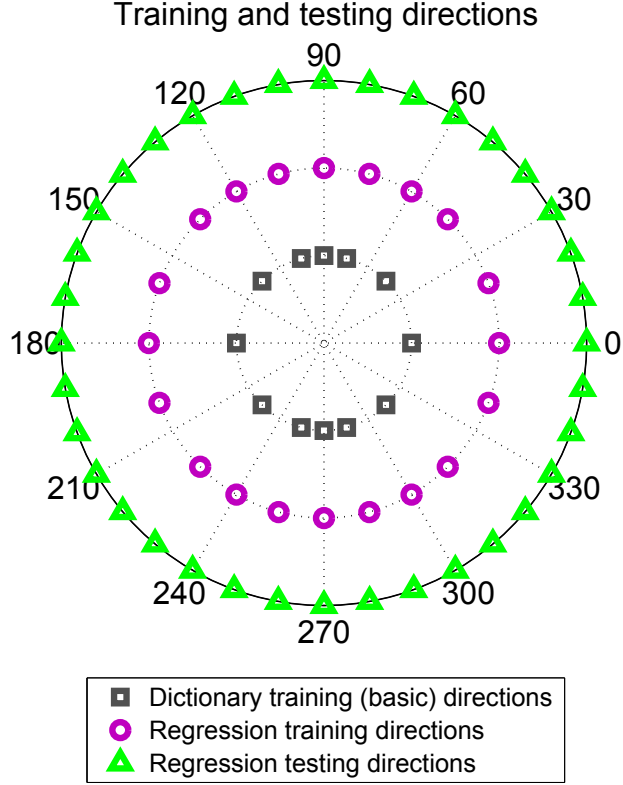}}
%\centerline{\psfig{figure=Fig8.eps,width=1.9in,height=2.4in}}
\caption{Schematic representation of the dictionary training, regression training and regression testing angles.} \label{anglesfig}
\end{figure}

The scenarios with $C=12$ basic directions of motion, $C_t=20$ regression training, and $C_s = 36$ regression testing directions were simulated. Fig. \ref{anglesfig} summarizes the selected directions of the pedestrian motion. Notice, the MD signature of the walking pedestrian becomes weaker as the direction of motion approaches the endfire region around $90^{\circ}$ or $270^{\circ}$. Fig. \ref{anglesfig} shows that the basic directions chosen for the dictionary training, and the regression training directions have a nonuniform spacing. Selection of more dense samples of direction of motion in the endfire region provides more training data and therefore compensates for weaker MD signatures. In addition, in order to keep the feature space dimensionality low, the number of dictionary training angles was selected to be $C < C_t$. In total $T_{tot} = 60$ seconds of the radar slow time data were generated for each direction of motion in the dictionary, and for the regression algorithm training and testing. The parameters of the SVR and the MLP regression were estimated using a 2-fold cross-validation. Table \ref{prmtable} summarizes the simulation parameters.

\begin{table}[!t]
\caption{List of the simulation parameters.}
\centering
\begin{tabular}{ c | p{3.5cm} | p{2.5cm} }
\hline
Parameter & Description & Value \\ \hline
$C$  & Number of basic dictionary training angles/Feature dimensionality & $12$\\ \hline
$C_t$  & Number of regression training angles & $20$\\ \hline
$C_s$  & Number of regression testing angles & $36$\\ \hline
$N$  & Number of spatial cells & $8$ \\ \hline
$T_{tot}$ & Length of the available radar slow-time database in seconds& 60\\ \hline
$X$  & Length of the dictionary training data for one orientation in samples & $60000$ \\ \hline
$K$  & Frame size & $32$ \\ \hline
$U$  & Number of columns in $\Ymat^c$ & $2 \lfloor X/K \rfloor - 1$, for $50\%$ overlap between frames  \\ \hline
$k$ & Number of columns in the dictionaries $\Dmat^c$ & 750 \\ \hline
$\xi$ & Regularization coefficient & 0.13 \\ \hline
$F_t$ & Number of data blocks available for regression training & $T_{tot} / T_F$ \\ \hline
$X_t$  & Regression training data block size (number of transmitted pulses in $T_F$ seconds) & $T_F / T_r$ \\ \hline
$U_t$  & Number of columns in $\Ymat_f^{c_t}$ & $2 \lfloor X_t/K \rfloor- 1$, for $50\%$ overlap between frames \\ \hline
$F_s$ & Number of data blocks available for regression testing & $F_t$ \\ \hline
\end{tabular}
\label{prmtable}
\end{table}

\subsection{Regression Error}
\label{RegErrSubSec}
This subsection presents simulation results in a scenario with the MIMO radar configuration with $L_{ty} = 4$ by $L_{tz} = 1$ transmitting and $L_{ry} = 4$ by $L_{rz} = 3$ receiving arrays discussed in Section \ref{subsec:AzimElev}, and orthogonal LFM waveforms with parameters discussed in Section \ref{subsec:RangeDoppler}.

\begin{figure}[htb]
\centerline{\includegraphics[scale=0.5]{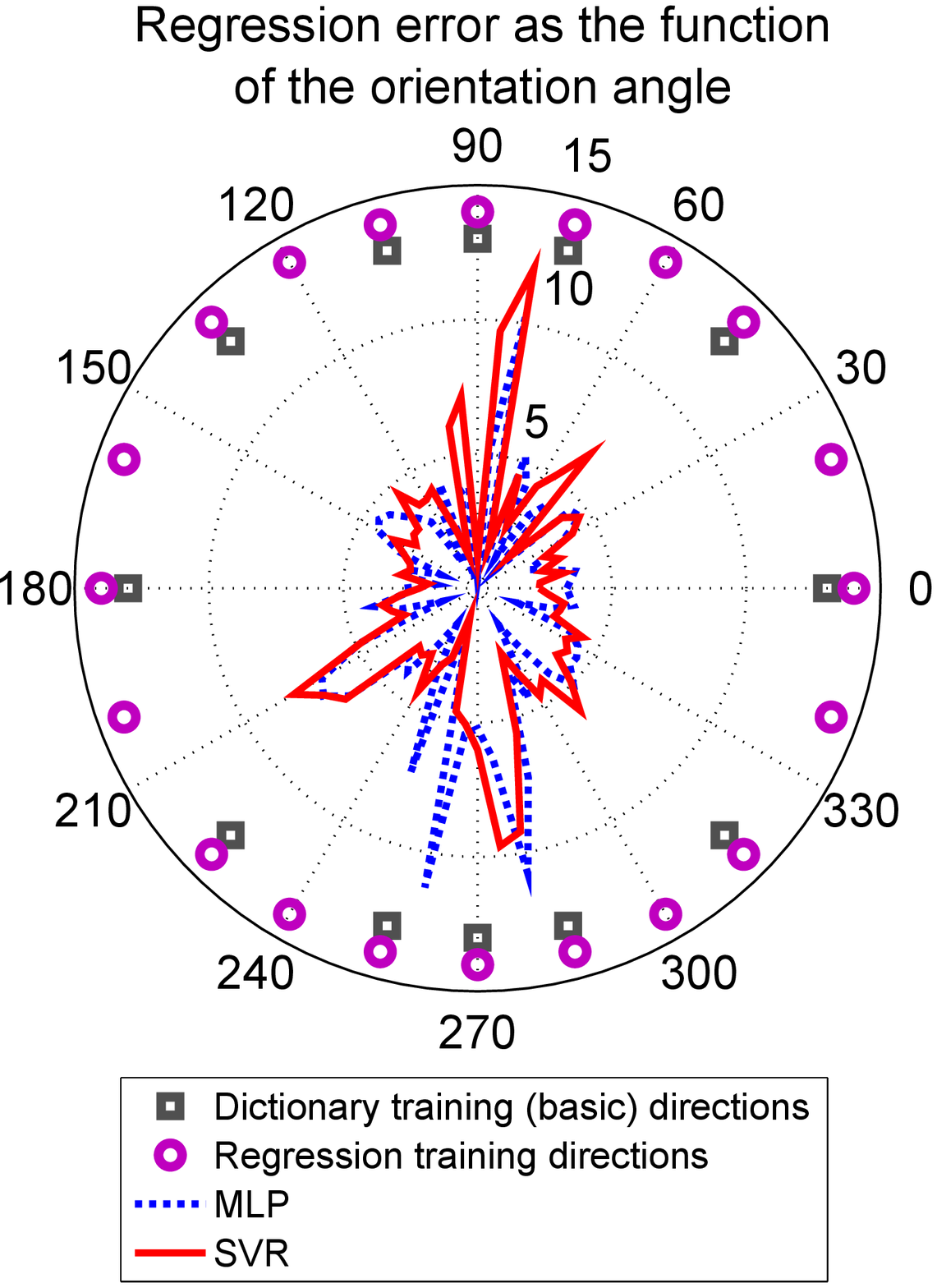}}
%\centerline{\psfig{figure=Fig9.eps,width=2in,height=2.6in}}
\caption{The regression error $\varepsilon^{c_s}$ as a function of the target direction of motion $\theta^{c_s}$ for the SVR and the MLP methods. SNR = 15dB, $T_F = 1$sec. The results for each SNR are averaged over $F_s = 60$ trials.} \label{polarfig}
\end{figure}

The regression performance was evaluated using a regression error criterion, defined as a square root of the MSE in (\ref{MSE}) as $\varepsilon^{c_s} = \left[MSE^{c_s}\right]^{1/2}$. The regression error averaged over $F_s = 60$ trials for SNR = 15dB and $T_F = 1$sec is shown in Fig. \ref{polarfig} as a function of the direction of motion. Fig. \ref{polarfig} shows that both the SVR and the MLP have larger errors at the angles that are not represented in the training sets, and that the error increases at the angles close to $90^{\circ}$ and $270^{\circ}$. These directions correspond to the scenarios where pedestrian walks perpendicular to the radar boresight and as the result produces a weaker MD signature. This limitation can be resolved by using larger number of training angles in the expense of increased size of the data sets, feature dimensionality and as the result the computation complexity and latency. Fig. \ref{polarfig} shows that the SVR outperforms the MLP with the regression error less than $5^{\circ}$ for a majority of tested motion directions. Notice that the regression error is extremely low in the scenarios with pedestrian moving directly towards or away from the radar. Such an accurate estimation performance is important for automotive active safety features, such as collision avoidance systems.

\begin{figure}[htb]
\centerline{\includegraphics[scale=0.5]{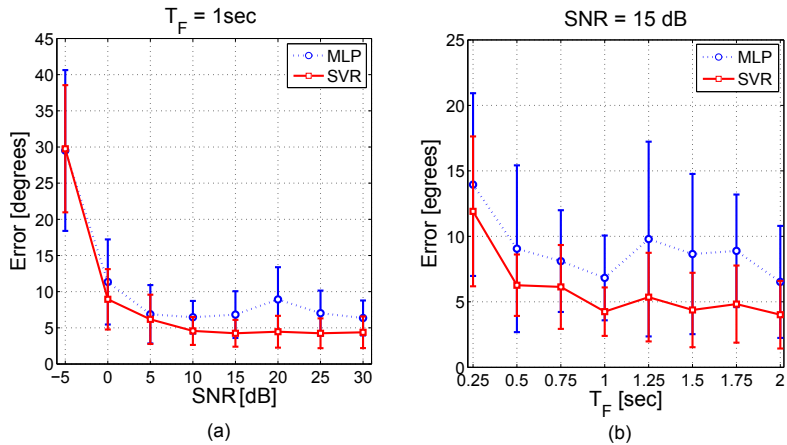}}
%\centerline{\psfig{figure=Fig10.eps,width=3.5in,height=2.0in}}
\caption{The average regression error $\varepsilon$ for the SVR and the MLP methods for a MIMO radar with the rectangular antenna array (8 spatial cells): (a) as a function of the SNR, $T_F=1$sec; (b) as a function of the target observation time $T_F$, SNR = 15dB. The results for each SNR and $T_F$ are averaged over all testing directions and $T_{tot}/T_F$ trials.} \label{snrtfig}
\end{figure}

Fig. \ref{snrtfig}(a) shows the averaged (over directions) regression error $\varepsilon = (1/C_s)\sum_{c_s}^{C_s}\varepsilon^{c_s}$, of the SVR and the MLP as a function of SNR. The target observation time was selected to be $T_F = 1$sec for all simulated SNR values. The minimum separation between the two adjacent training angles was $15^{\circ}$. The error bars indicate the standard deviation of the error. Fig. \ref{snrtfig}(a) shows that both methods demonstrate a good performance for the SNR higher than $10$dB, and that the SVR slightly outperforms the MLP, providing error less than $5^{\circ}$ with the standard deviation less than $2^{\circ}$. Notice that the simulation results in Fig. \ref{snrtfig} were obtained using the MLP network with fixed weights, and their optimization is expected to improve MLP performance.

\begin{figure}[htb]
\centerline{\includegraphics[scale=0.5]{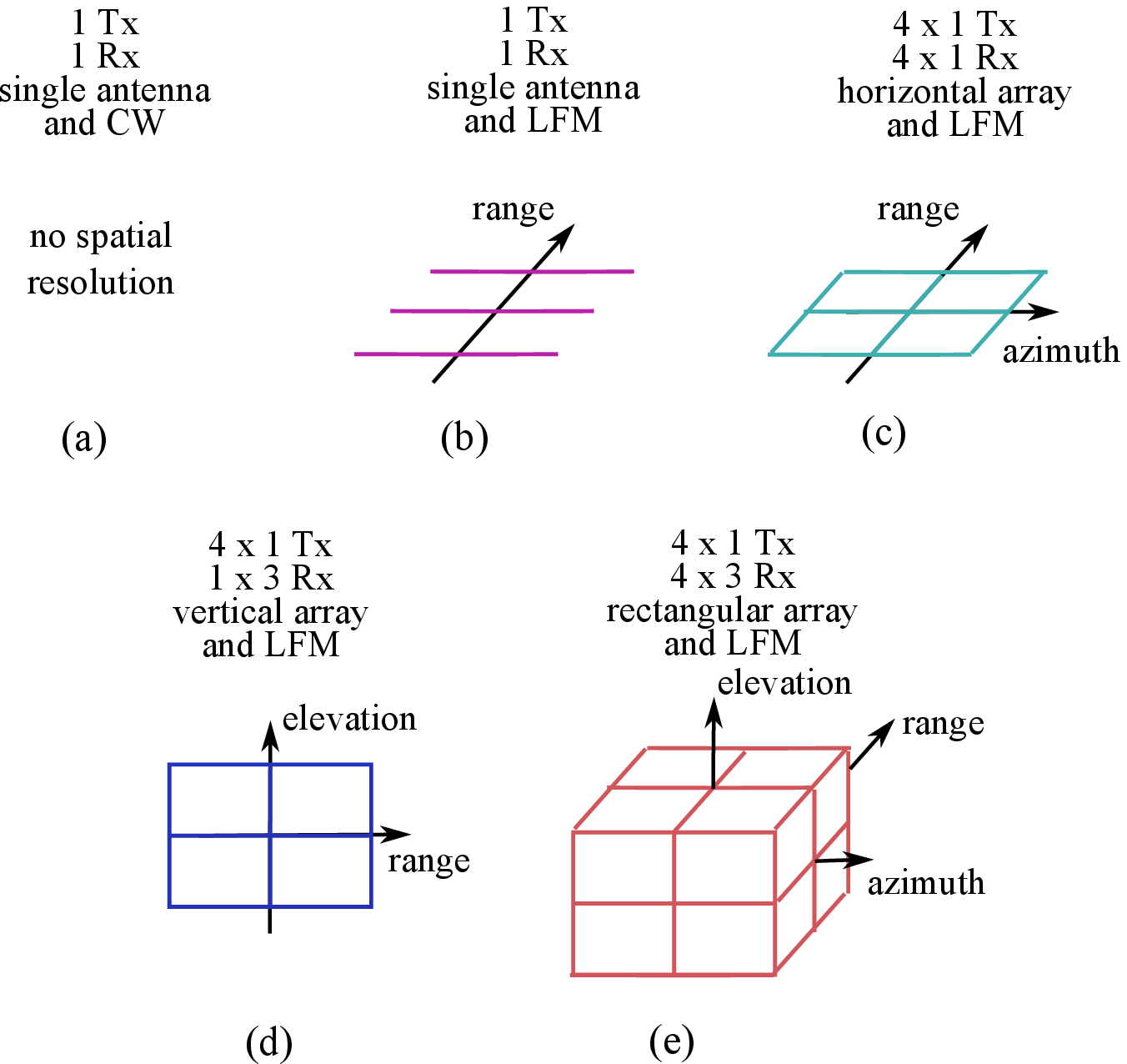}}
%\centerline{\psfig{figure=Fig11.eps,width=2.8in,height=2.3in}}
\caption{Different spatial grids and corresponding radar configurations.} \label{cells}
\end{figure}

Fig. \ref{snrtfig}(b) shows the estimation error as the function of the target observation time $T_F$ for the fixed SNR of $15$dB. Notice that the average regression error is lowest when the target observation time is higher than $T_F = 1$sec. High regression errors for $T_F < 1$ occur since the target observation time interval is too short to obtain sufficient information about the target direction of motion.

The rest of the section presents the simulation results for the SVR regression method only, since it outperforms the MLP in all tested scenarios.

\subsection{Radar Configuration Comparison}
This subsection investigates the influence of the radar configuration and the corresponding spatial grid on the motion direction estimation. The following five radar configurations that result in different spatial grids shown in Fig. \ref{cells} (the spatial grids shown in Fig. \ref{cells} consist only of the relevant cells, i.e. the cells which contain the target) are compared:

\begin{enumerate}
\item The radar with a single-element antenna that transmits a continuous wave (CW): such radar has no spatial resolution.

\item The radar with a single-element antenna which transmits the LFM waveform with parameters discussed in Section \ref{subsec:RangeDoppler}. Such radar configuration provides the range resolution i.e. the spatial grid consists of the two cells that are placed along the range dimension as shown in Fig. \ref{cells}(b).

\item A MIMO radar with a $4 \times 1$ transmitting array discussed in Section \ref{subsec:AzimElev}, and with a $4 \times 1$ horizontal array of receiving elements. The beam pattern of such MIMO radar is equal to the horizontal cut through the zero elevation line of the 2D beam pattern shown in Fig. \ref{Fig3}(b). Such configuration provides the azimuth and the range resolution, but has no elevation resolution. The spatial grid consists of the four cells located in the horizontal plane (Fig. \ref{cells}(c)). 

\item A MIMO radar with a $4 \times 1$ transmitting array discussed in Section \ref{subsec:AzimElev}, and with a $1 \times 3$ vertical receiving array. The beam pattern of this MIMO radar is equal to the vertical cut through the zero azimuth line of the 2D beam pattern shown in Fig. \ref{Fig3}(b). Such a MIMO radar configuration provides the elevation and the range resolution, but has no azimuth resolution. The corresponding spatial grid consists of the four cells located in the vertical plane (Fig. \ref{cells}(d)).

\item A MIMO radar with a $4 \times 1$ transmitting and a $4 \times 3$ receiving arrays discussed in Section \ref{RegErrSubSec}. The spatial grid consists of the 8 spatial cells (Fig. \ref{cells}(e)).
\end{enumerate}

\begin{figure}[htb]
\centerline{\includegraphics[scale=0.5]{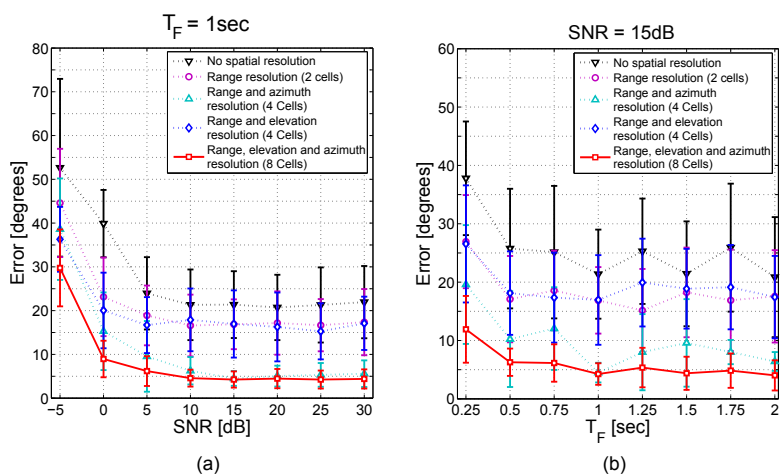}}
%\centerline{\psfig{figure=Fig12.eps,width=3.5in,height=2.0in}}
\caption{The average regression error $\varepsilon$ for the SVR method and different radar configurations: (a) show the regression error as the function of the SNR, and $T_F = 1$sec; (b) show the regression error as the function of the target observation time $T_F$, and SNR=15dB.} \label{allres}
\end{figure}

The pedestrian direction of motion estimation performance for the considered radar configurations 1-5 and SVR regression method is summarized in Fig. \ref{allres}. Notice that the configuration 3 with a horizontal receiving array and configuration 4 with a vertical receiving array have equal number of spatial cells, however the horizontal receiving array provides a significantly smaller regression error than the vertical receiving array. Therefore, the configuration in Fig. \ref{cells}(c) with the spatial cells located in the horizontal plane is more beneficial for the problem of direction of motion estimation than the configuration in Fig. \ref{cells}(d) with the spatial cells located in the vertical plane. This effect can be explained by the fact that the pedestrian's body parts perform mostly horizontal motions, and their relative locations can be resolved in the horizontal plane.

This preliminary analysis demonstrates a possibility to improve the pedestrian direction of motion estimation by proper selection of MIMO radar architecture, and the optimal MIMO radar architecture is the subject of our future research.

\begin{figure}[htb]
\centerline{\includegraphics[scale=0.5]{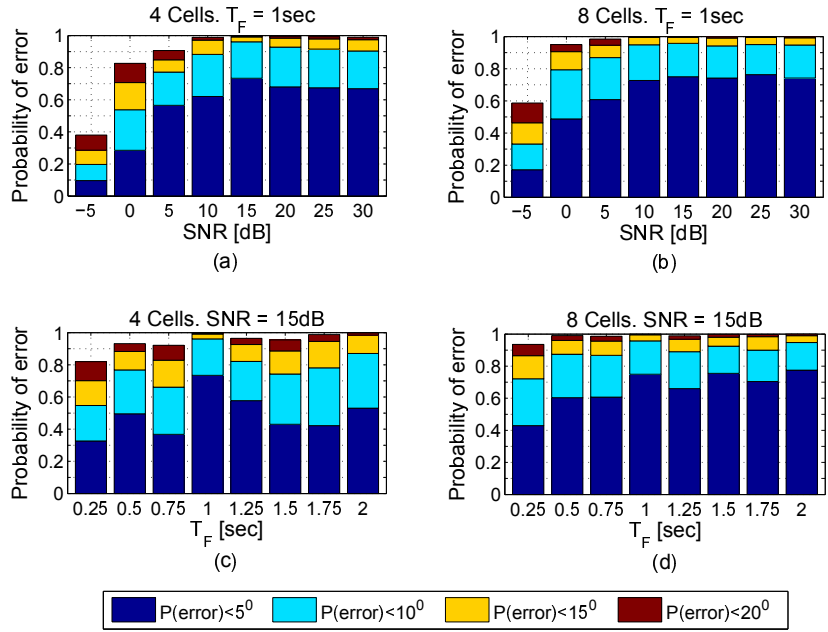}}
%\centerline{\psfig{figure=Fig13.eps,width=3.5in,height=2.5in}}
\caption{The probability of the target's motion direction estimation error to be lower than $5^{\circ}$, $10^{\circ}$, $15^{\circ}$ and $20^{\circ}$: (a) as a function of the SNR ($T_F = 1$s) for a radar configuration with the 4 horizontal spatial cells and no elevation resolution; (b) as a function of the SNR ($T_F = 1$s) for a radar configuration with the 8 spatial cells and the elevation resolution; (c) as a function of $T_F$ (SNR=15db) for a radar configuration with the 4 horizontal spatial cells and no elevation resolution; (d) as a function of $T_F$ (SNR=15db) for a radar configuration with the 8 spatial cells and the elevation resolution. The results are obtained using the SVR method.} \label{bars}
\end{figure}

\subsection{Probability of Error}
Fig. \ref{allres} shows that the MIMO radar with a rectangular receiving array and with a horizontal receiving array have comparable performance. This subsection further investigates the influence of the elevation resolution on the pedestrian direction of motion estimation performance via evaluation of the probabilities of the average regression error being less than $5^{\circ}$, $10^{\circ}$, $15^{\circ}$ and $20^{\circ}$ (a percent of the test frames which have direction estimation error smaller than a given value) for both MIMO radar configurations.

The averaged probabilities of the regression error for the MIMO radar with the horizontal receiving array and the MIMO radar with the rectangular receiving array are show in Fig. \ref{bars}(a) and Fig. \ref{bars}(c), and Fig. \ref{bars}(b) and Fig. \ref{bars}(d). respectively. Fig. \ref{bars}(a) and Fig. \ref{bars}(b) show that for the SNR=15dB and $T_F = 1$sec, the probability of the regression error being less than $10^{\circ}$ is $0.95$ for both MIMO radar configurations. Therefore, for the SNR above $15$dB the elevation resolution does not provide significant improvement in motion direction estimation, and good direction estimation results can be obtained with the horizontal antenna array only. However, in the low-SNR scenarios, the MIMO radar with the 2D receiving array has a better performance at the expense of 3 times larger number of receiving antenna elements.

The performance of the proposed supervised learning-based approach is heavily dependent on the quantity and quality of the available training data. In the presented simulation results we assumed that 60 seconds of the radar slow time data are available for each training and testing direction. A smaller training data set would result in a degraded performance and larger estimation errors. Furthermore, the considered in this paper radar signal model does not take into account a number of real world effects such as reflections from the road surface, surrounding buildings, vegetation, other vehicles and pedestrians, as well as the influence of the weather conditions, such as rain and snow, on the radar signal propagation. These phenomena will significantly affect the quality of the training data in a practical automotive radar system. Hence, a sufficient amount of training data needs to be collected in various scenarios and weather conditions such that different propagation and reflection effects are well represented in the training data set. In addition, the training data set needs to include a diverse population of pedestrians which have different heights and walk with different velocities in order to guarantee high generalization capabilities of the trained regression model.

\section{Conclusions}
This work proposed a regression-based method for pedestrian direction of motion estimation using its MD signatures obtained by the automotive MIMO radar. Performance of the SVR and the MLP regression methods was evaluated via simulations as a function of the SNR, observation time and MIMO radar configuration. It was shown that a good direction of motion estimation performance (with error less than $5^{\circ}$) can be achieved using the SVR-based method in majority of tested directions of motion. It was also shown that the estimation performance improves for motion directions toward the radar, and degrades for motion angles perpendicular to the radar boresight.

Considering various MIMO radar configurations it was shown that the direction of motion estimation performance improves with increasing the number of horizontal array elements (higher azimuth resolution). Finally it was shown that for the low-SNR scenarios vertical resolution also improves the direction of motion estimation performance.

%\bibliographystyle{plain}
%\bibliography{GeneralBoundForRadarBibliographyFile}

\begin{thebibliography}{99}
\bibitem{Buehler} Buehler, M. {\em The DARPA Urban Challenge: Autonomous Vehicles in City Traffic, 1st ed.}, New York: Springer-Verlag, 2009.
\bibitem{Urmson1} Urmson, C. and Whittaker, W. ``Self-driving cars and the urban challenge,'' {\em IEEE Intelligent Systems}, vol. 23, no. 2, pp. 66-68, 2008.
\bibitem{Urmson2} Urmson, C., et al. ``Autonomous driving in urban environments: Boss and the urban challenge,'' {\em Journal of Field Robotics} vol. 25, no. 8, pp. 425-466, 2008.
\bibitem{Luettel} Luettel T., Himmelsbach M., and Wuensche, H. ``Autonomous ground vehicles - concepts and a path to the future,'' {\em In proc. of the IEEE, Special Centennial Issue}, vol.100, pp.1831-1839, 2012.
\bibitem{Levinson} Levinson, J., et al., ``Towards fully autonomous driving: Systems and algorithms,'' {\em In Proc. of IEEE  Intelligent Vehicles Symposium}, pp. 163-168, 2011.
\bibitem{Fleming} Fleming, W. ``Overview of automotive sensors,'' {\em IEEE Sensors Journal}, vol. 1, no. 4, pp. 296-308, 2001.
\bibitem{Rohling} Rohling, H. ``Milestones in radar and the success story of automotive radar systems,'' {\em In Proc. of 11th IEEE International Radar Symposium}, 2010.
\bibitem{Murad} Murad, M., et al., ``Requirements for next generation automotive radars,'' {\em In Proc. of IEEE Radar Conference}, 2013.
\bibitem{Heuel1} Heuel, S. and Rohling, H. ``Pedestrian classification in automotive radar systems,'' {\em In Proc. of the 19th International Radar Symposium (IRS),} pp. 39-44, May, 2012.
\bibitem{Heuel2} Rohling, H., Heuel, S., and Ritter, H. ``Pedestrian detection procedure integrated into an 24 GHz automotive radar,'' {\em In proc. of the IEE Radar Conference,} pp.1229-1232, 2010.
\bibitem{Heuel3} Heuel, S. and Rohling, H. ``Two-stage pedestrian classification in automotive radar systems,'' {\em In Proc. of the 2011 International Radar Symposium (IRS),}. pp. 477-484, 2011.
\bibitem{Cortes} Cortes, C. and Vapnik, V. ``Support-vector networks,'' {\em Machine learning} vol. 20, no. 3, pp. 273-297, 1995.
\bibitem{DoganNehor1} Dogandzic, A. and Nehorai, A. ``CramerRao bounds for estimating range, velocity, and direction with an active array," {\em IEEE Trans. Signal Process.,} vol. 49, no. 6, pp. 1122-1137, 2001.
\bibitem{LiStoica1} Li, J. and Stoica, P. ``MIMO radar with colocated antennas," {\em IEEE Signal Process. Mag.,} vol. 24, no. 5, pp. 106-114, 2007.
\bibitem{LiStoica0} Li, J. and Stoica, P. {\em MIMO Radar Signal Processing.} Hoboken, NJ: Wiley, 2009.
\bibitem{Skolnik} Skolnik, M. {\em Introduction to radar systems.} New York: McGraw-hill, 2001.
\bibitem{Richards2} Richards, M. {\em Fundamentals of radar signal processing.} Tata McGraw-Hill Education, 2005.
\bibitem{Bellman} Bellman, R. {\em Adaptive control processes: a guided tour.} Princeton: Princeton University Press, 1961.
\bibitem{Sapiro1} Castrodad, A. and Sapiro, G. ``Sparse modeling of human actions from motion imagery," {\em Springer, International Journal of Computer Vision,} vol. 100, no. 1, pp. 1-15, 2012.
\bibitem{Sapiro09} Mairal, J., Bach, F., Ponce, J., and Sapiro, G. ``Online dictionary learning for sparse coding," {\em In Proc. of the 26th Annual International Conference on Machine Learning}, pp. 689-696, 2009.
\bibitem{Sapiro10} Mairal, J., Bach, F., Ponce, J., and Sapiro, G. ``Online learning for matrix factorization and sparse coding," {\em The Journal of Machine Learning Research} vol. 11, pp. 19-60, 2010.
\bibitem{Bradley} Bradley, M. ``Human Walking Animation" from the Wolfram Demonstrations Project http://demonstrations.wolfram.com/HumanWalkingAnimation/
\bibitem{Chen1} Chen, V., Li, F., Ho, S., Wechsler, H. ``Micro-Doppler effect in radar: phenomenon, model, and simulation study," {\em IEEE Trans. on Aerospace and Electronic Systems,} vol. 42, no. 1, pp. 2-21, 2006.
\bibitem{Chen11} Chen, V. The micro-Doppler effect in radar., {\em Artech House Publishers,} 2011.
\bibitem{Chen2} Chen, V. ``Doppler signatures of radar backscattering from objects with micro-motions,'' {\em  IET Signal Processing}, vol. 2, no. 3, pp. 291-300, 2008.
\bibitem{Boulic} Boulic, R., Thalmann, N., and Thalmann, D. ``A global human walking model with real-time kinematic personification," {\em Visual Computer,} vol. 6, no. 6, pp. 344-358, 1990.
\bibitem{Trott} Trott, K. ``Stationary phase derivation for RCS of an ellipsoid," {\em IEEE Antenna and Wireless Propagation Letters,} vol. 6, pp. 240-243, 2007.
\bibitem{KimLing09} Kim, Y. and Ling, H. ``Human activity classification based on micro-Doppler signatures using a support vector machine," {\em IEEE Trans. on Geoscience and Remote Sensing,} vol. 47, no. 5, pp. 1328-1337, 2009.
\bibitem{LeiLu05} Lei, J., and Lu, C. ``Target classification based on micro-Doppler signatures," {\em In Proc. of IEEE Radar Conference}, pp. 179-183, 2005.
\bibitem{Bar} Bar-Hillel, A., Bilik, I., Hecht, R. ``Naive Bayes nearest neighbor classification of ground moving targets," {\em In Proc. of IEEE Radar Conference}, 2013.
\bibitem{Tekeli} Tekeli, B., et al., ``Classification of human micro-Doppler in a radar network," {\em In Proc. of IEEE Radar Conference}, 2013.
\bibitem{SWB1} Smith, G., Woodbridge, K., and Baker, C. ``Micro-Doppler signature classification," {\em In Proc. of IEEE Radar Conference}, pp. 111-116, 2006.
\bibitem{SWB2} Smith, G., Woodbridge, K., and Baker, C. ``Radar micro-Doppler signature classification using dynamic time warping," {\em IEEE Trans. on Aerospace and Electronic Systems,} vol. 46, no. 3, pp. 1078-1096, 2010.
\bibitem{NanzerRogers09} Nanzer J. and  Rogers, R. ``Bayesian classification of humans and vehicles using Micro-Doppler signals from a scanning-beam radar," {\em IEEE Microwave and Wireless Components Letters}, vol. 19, no. 5, pp. 338-340, 2009.
\bibitem{Thayaparan08} Thayaparan, T., et al, ``Micro-Doppler-based target detection and feature extraction in indoor and outdoor environments," {\em Journal of the Franklin Institute} vol. 345, no. 6, pp. 700-722, 2008.
\bibitem{Bilik06} Bilik, I., Tabrikian, J., and Cohen, A. ``GMM-based target classification for ground surveillance Doppler radar," {\em IEEE Trans. on Aerospace and Electronic Systems}, vol. 42, no. 1, pp. 267-278, 2006.
\bibitem{Bilik12} Bilik, I., and Khomchuk, P. ``Minimum divergence approaches for robust classification of ground moving targets," {\em IEEE Trans. on Aerospace and Electronic Systems}, vol. 48, no. 1, pp. 581-603, 2012.
\bibitem{Bilik07} Bilik, I. and Tabrikian, J. ``Radar target classification using Doppler signatures of human locomotion models," {\em IEEE Trans. on Aerospace and Electronic Systems}, vol. 43, no. 4, pp. 1510-1522, 2007.
\bibitem{Duda} Duda, R., Hart, R., and Stork, D. Pattern Classification., {\em Hoboken, NJ: Wiley,\/} 2001.
\bibitem{Smola97} Drucker, H., Burges, C., Kaufman, L., Smola, A., and Vapnik, V. ``Support vector regression machines. In: Mozer M., Jordan M., and Petsche T. (Eds.)," {\em Advances in neural information processing systems 9,} MIT Press, Cambridge, MA, pp. 155-161, 1997.
\bibitem{Smola04} Smola, A., and Schölkopf, B. ``A tutorial on support vector regression," {\em Statistics and computing,} vol. 14, no. 3, pp. 199-222, 2004.
\bibitem{Hornik89} Hornik, K., Stinchcombe, M. and White, H. ``Multilayer feedforward networks are universal approximators," {\em Neural networks } vol. 2, no. 5, pp. 359-366, 1989.
\bibitem{Fairchild} Fairchild, D., and Narayanan, R. ``Determining human target facing orientation using bistatic radar micro-Doppler signals," {\em In Proc. of SPIE Conference on Active and Passive Signatures V}, vol. 9082, pp. 1-9, 2014.
\bibitem{Suykens} Suykens, J., et al., {\em Least Squares Support Vector Machines.} World Scientific, Singapore, 2002.
\bibitem{Nabney} Nabney, I. {\em NETLAB: Algorithms for pattern recognition.} Springer, 2002.
\end{thebibliography}
\end{document}